\documentclass[fleqn,usenatbib,useAMS]{mnras}  

\usepackage{newtxtext,newtxmath}  
\usepackage[T1]{fontenc}  
\usepackage{amsmath}  
\usepackage{graphicx}
\usepackage{xspace}
\usepackage{orcidlink}

\title[ACES Filaments in the CMZ]{ALMA Central molecular zone Exploration Survey (ACES) VI: ALMA Large Program Reveals a Highly Filamentary Central Molecular Zone}

\newcounter{affcounter}

\newcommand{\defaffiliationlabel}[1]{%
  \refstepcounter{affcounter}%
  \expandafter\xdef\csname #1\endcsname{\theaffcounter}%
}

\newcommand{\affref}[1]{$^{\csname #1\endcsname}$}

\newcommand{\affrefs}[1]{%
  $^{%
    \@for\@ref:=#1\do{%
      \@ref\@ifnextchar\@nil{}{,}%
    }%
  }$%
}

\newcommand{\affrefTwo}[2]{$^{\csname #1\endcsname,\csname #2\endcsname}$}
\newcommand{\affrefThree}[3]{$^{\csname #1\endcsname,\csname #2\endcsname,\csname #3\endcsname}$}
\newcommand{\affrefFour}[4]{$^{\csname #1\endcsname,\csname #2\endcsname,\csname #3\endcsname,\csname #4\endcsname}$}

\newcommand{\printaffiliation}[2]{%
  $^{\csname #1\endcsname}$#2\\%
}
\defaffiliationlabel{uconn}
\defaffiliationlabel{uflorida}
\defaffiliationlabel{iff}
\defaffiliationlabel{villanova}
\defaffiliationlabel{cassaca}
\defaffiliationlabel{ucn}
\defaffiliationlabel{cab_csic}
\defaffiliationlabel{nanjing}
\defaffiliationlabel{nanjing_key}
\defaffiliationlabel{cfa}
\defaffiliationlabel{ukarcnode}
\defaffiliationlabel{jpl}
\defaffiliationlabel{ias_hubble}
\defaffiliationlabel{uw_madison}
\defaffiliationlabel{shao}
\defaffiliationlabel{naoc_key}
\defaffiliationlabel{kansas}
\defaffiliationlabel{eso}
\defaffiliationlabel{naoj}
\defaffiliationlabel{ljmu}
\defaffiliationlabel{mpia}
\defaffiliationlabel{cool}
\defaffiliationlabel{colorado}
\defaffiliationlabel{iaa_taipei}
\defaffiliationlabel{eao}
\defaffiliationlabel{chalmers}
\defaffiliationlabel{surrey}
\defaffiliationlabel{insubria}
\defaffiliationlabel{nrao}
\defaffiliationlabel{ita_heidelberg}
\defaffiliationlabel{northwestern_ciera}
\defaffiliationlabel{mpir}
\defaffiliationlabel{ucl}
\defaffiliationlabel{izw_heidelberg}
\defaffiliationlabel{radcliffe}
\defaffiliationlabel{eso_chile}
\defaffiliationlabel{jao}
\defaffiliationlabel{mpe}
\defaffiliationlabel{ulaserena_postgrad}
\defaffiliationlabel{ulaserena}
\defaffiliationlabel{ice_csic}
\defaffiliationlabel{ieec}
\defaffiliationlabel{gbo}
\defaffiliationlabel{utokyo}
\defaffiliationlabel{iaa_csic}
\defaffiliationlabel{manchester}

\author[C. Battersby et al.]{Cara Battersby,\affref{uconn}\orcidlink{0000-0002-6073-9320}
Miriam G. Santa-Maria,\affrefTwo{uflorida}{iff}\orcidlink{0000-0002-3941-0360}
Dani Lipman,\affref{uconn}\orcidlink{0000-0002-5776-9473}
Dylan M. Par\'e,\affref{villanova}\orcidlink{0000-0002-5811-0136}
Rachel R. Lee,\affref{uconn}\orcidlink{0000-0002-7482-5078}
\newauthor
Pablo Garc\'ia,\affrefTwo{cassaca}{ucn}\orcidlink{0000-0002-8586-6721}
Izaskun Jim\'enez-Serra,\affref{cab_csic}\orcidlink{0000-0003-4493-8714}
Xing Pan,\affrefThree{nanjing}{nanjing_key}{cfa}\orcidlink{0000-0003-1337-9059}
Daniel L. Walker,\affref{ukarcnode}\orcidlink{0000-0001-7330-8856}
Jack Sullivan,\affref{uconn}
\newauthor
Danya Alboslani,\affref{uconn}\orcidlink{0009-0005-9578-2192}
H Perry Hatchfield,\affrefTwo{jpl}{uconn}\orcidlink{0000-0003-0946-4365}
Yue Hu,\affref{ias_hubble}\orcidlink{0000-0002-8455-0805}
Alex Lazarian,\affref{uw_madison}
Jennifer Wallace,\affref{uconn}\orcidlink{0009-0002-7459-4174}
\newauthor
Qizhou Zhang,\affref{cfa}\orcidlink{0000-0003-2384-6589}
Xing Lu,\affrefTwo{shao}{naoc_key}\orcidlink{0000-0003-2619-9305}
Elisabeth A.C. Mills,\affref{kansas}\orcidlink{0000-0001-8782-1992}
Adam Ginsburg,\affref{uflorida}\orcidlink{0000-0001-6431-9633}
Ashley~T.~Barnes,\affref{eso}\orcidlink{0000-0003-0410-4504}
\newauthor
Pei-Ying Hsieh,\affref{naoj}\orcidlink{0000-0001-9155-3978}
Jonathan D. Henshaw,\affrefTwo{ljmu}{mpia}\orcidlink{0000-0001-9656-7682}
Steven N. Longmore,\affrefTwo{ljmu}{cool}\orcidlink{0000-0001-6353-0170}
John Bally,\affref{colorado}\orcidlink{0000-0001-8135-6612}
Laura Colzi,\affref{cab_csic}\orcidlink{0000-0001-8064-6394}
\newauthor
Paul T. P. Ho,\affrefTwo{iaa_taipei}{eao}\orcidlink{0000-0002-3412-4306}
Maya A. Petkova,\affref{chalmers}\orcidlink{0000-0002-6362-8159}
Mattia C. Sormani,\affrefTwo{surrey}{insubria}\orcidlink{0000-0001-6113-6241}
N. Bijas,\affref{manchester}\orcidlink{0000-0002-6398-7530}
Alyssa Bulatek,\affref{uflorida}\orcidlink{0000-0002-4407-885X}
\newauthor
Natalie O. Butterfield,\affref{nrao}\orcidlink{0000-0002-4013-6469}
Christoph Federrath,\affref{anu}\orcidlink{0000-0002-0706-2306}
Simon C.~O.\ Glover,\affref{ita_heidelberg}\orcidlink{0000-0001-6708-1317}
Mark D.\ Gorski,\affref{northwestern_ciera}\orcidlink{0000-0001-9300-354X}
\newauthor
Savannah R. Gramze,\affref{uflorida}\orcidlink{0000-0002-1313-429X}
Christian Henkel,\affref{mpir}\orcidlink{0000-0002-7495-4005}
Janik Karoly,\affref{ucl}\orcidlink{0000-0001-5996-3600}
Ralf S.\ Klessen,\affrefFour{ita_heidelberg}{izw_heidelberg}{cfa}{radcliffe}\orcidlink{0000-0002-0560-3172}
Sergio Mart\'in,\affrefTwo{eso_chile}{jao}\orcidlink{0000-0001-9281-2919}
\newauthor
Francisco Nogueras-Lara,\affrefTwo{iaa_csic}{eso}\orcidlink{0000-0002-6379-7593}
Jaime E. Pineda,\affref{mpe}\orcidlink{0000-0002-3972-1978}
Denise Riquelme-V\'asquez,\affref{ulaserena}\orcidlink{0000-0001-5389-0535}
V\'ictor M. Rivilla,\affref{cab_csic}\orcidlink{0000-0002-2887-5859}
\newauthor
\'Alvaro S\'anchez-Monge,\affrefTwo{ice_csic}{ieec}\orcidlink{0000-0002-3078-9482}
Anika Schmiedeke,\affref{gbo}\orcidlink{0000-0002-1730-8832}
Yoshiaki Sofue,\affref{utokyo}\orcidlink{0000-0002-4268-6499}
and Volker Tolls\affref{cfa}\orcidlink{0000-0003-1841-2241}
\\
$^{*}$Author affiliations are listed at the end of the paper
}

\date{Accepted XXX. Received YYY; in original form ZZZ}

\pubyear{2025}

\begin{document}
\label{firstpage}
\pagerange{\pageref{firstpage}--\pageref{lastpage}}
\maketitle

\newcommand\HNCO{HNCO\;\mbox{4(0,4)--3(0,3)}}
\newcommand\SiO{\mbox{SiO\;(2--1)}}
\newcommand\CS{\mbox{CS\;(2--1)}}
\newcommand\SO{\mbox{SO\;3(2)--2(1)}}
\newcommand\HCtrN{\mbox{HC$_{3}$N\;(11--10)}}
\newcommand\HthCOp{\mbox{H$^{13}$CO$^{+}$\;(1--0)}}
\newcommand\HthCN{\mbox{H$^{13}$CN\;(1--0)}}
\newcommand\HNthC{HN$^{13}$C\;(1--0)}
\newcommand{\degree}{$^{\circ}$}
\newcommand{\kms}{\,km\,s$^{-1}$\xspace}
\newcommand{\hcop}{HCO$^{+}$\xspace}

\newcommand{\updates}[1]{\textcolor{black}{#1}}

\newcommand{\updab}[1]{\textcolor{black}{#1}}

\newcommand{\updmay}[1]{\textcolor{black}{#1}}

\begin{abstract}
The Central Molecular Zone (CMZ) of the Milky Way is the way station that primarily controls how much gas flows from the disk of the Galaxy towards the central nucleus. While this region is well documented to have extreme gas properties that clearly distinguish it from the rest of the Galaxy, the properties of the bulk molecular gas at high angular resolution are relatively unexplored. Band 3 data from the ALMA (Atacama Large Millimeter/Submillimeter Array) large program ACES (ALMA CMZ Exploration Survey) \updates{reveal} the highly filamentary nature of CMZ molecular gas at high resolution (3\arcsec~or $\sim$0.1~pc) across the entire CMZ.
Visual inspection of these data \updates{suggests} that there are at least two general classes of elongated structures, which we identify as: i) large-scale ($\gtrsim 10$ pc) filamentary structures (LFs) and ii) a ubiquitous population of small-scale ($\sim 1$ pc) filamentary structures (SFs)\updab{, both with widths (FWHM) of about 0.1 pc}. We present detailed morphological and kinematic properties towards three structures in each category, as well as their association with magnetic fields and the correlation of \HNCO~with other molecular species.
Our investigation reveals that these structures are largely coherent in position-position-velocity space. The alignment with the magnetic field structure is mixed, with some parallel, some perpendicular, and some intermediate alignments.
We find that LFs likely trace pieces of contiguous CMZ orbital structures and are a manifestation of global CMZ dynamics. The second class, SFs, are \updates{pervasive} and \updates{may be} the result of complicated turbulence and shearing dynamics in the CMZ gas flows, \updates{as seen in numerical simulations.}
\end{abstract}

\begin{keywords}
ISM: general -- ISM: molecules -- submillimetre: ISM -- Galaxy: centre -- Stars: formation
\end{keywords}

\section{Introduction} \label{sec:intro}

We live in a barred spiral galaxy, with gas properties that vary across the Galactic disk \citep[e.g.][]{Sandstrom2013, RomanDuval2014, Heyer2015, Ragan2016}. The Galactic bar drives material inwards along `dust lanes’ \citep[e.g.][]{Sormani2019, Hatchfield2021} to the Central Molecular Zone (CMZ). The CMZ is the concentration of dense molecular gas in the inner 400~pc of the Galaxy \citep{Henshaw2023, Battersby2025a}. Some of this gas forms stars or finds its way into the circumnuclear disk (CND) and the supermassive black hole SgrA* \citep[e.g.][]{2012ApJ...751..124K,Tress2020, Sormani2020, Lu2021,2024ApJ...966..230K,2024MNRAS.527.9343L, Tress2024}. Perhaps due to the potential energy of the material inflowing along the Galactic bar into the CMZ, gas in this region has distinct and extreme physical properties compared with gas elsewhere in the Galactic disk. \citet{Henshaw2023} provide a review of this dynamic region. Key differences in the gas properties include its higher density, temperature, degree of turbulence, and magnetic field strength, all of which tend to be 1-2 orders of magnitude higher compared to the Galactic disk \citep[e.g.][]{Pillai2015, Mills2018b, Ginsburg2016, Kauffmann2017, Federrath2016, Butterfield2024, Colzi2024}.

Most of the molecular gas mass in the CMZ orbits the Galactic Center at a distance of about 100 - 200 pc in approximately elliptical x$_2$ orbits \citep[e.g.][]{Walker2025, Lipman2025, Tress2024, Tress2020, Kruijssen2015}. \updates{Images in the} far-infrared dust continuum \updates{highlight key dense molecular gas structures in the CMZ, and have been described as a figure-eight or infinity symbol in appearance \citep[e.g.][]{Molinari2011}. We call this the} twisted ellipse. The kinematic decomposition of dense molecular gas tracers reveals a complicated though somewhat contiguous orbital stream of gas in position-position-velocity (PPV) space \citep{Henshaw2016a}. The 3D structure of these gas flows is a topic of debate including the `injection points’ of the gas flow from the larger Galaxy and possible connections with the CND or SgrA* \citep[see review by][]{Henshaw2023}. The 3D CMZ paper series \citep{Battersby2025a, Battersby2025b, Walker2025, Lipman2025} finds that no present model is a perfect fit to the regions' complicated 3D structure. However \citet{Walker2025} present an elliptical x$_2$ model that is a reasonable match with existing data.

Gas in the dust lanes on its way to entering the CMZ has been observed to be highly filamentary \citep{Wallace2022, Butterfield2025}, and simulations find that clouds already have filamentary shapes due to shearing in the dust lanes when they enter the CMZ \citep{Tress2020}. Very thin filaments \updab{(widths of 0.07-0.14 pc, lengths of 1.2-1.8 pc)} were detected in absorption with HCO$^+$ against the CMZ molecular cloud, known as the ``Brick,” using ALMA data in \citet{Bally2014}. \citet{Yang2025} identify slim \updates{($\lesssim  0.03$ pc)} filaments in the CMZ tracing pc-scale shocks. However, the preponderance of filaments and their properties in CMZ molecular gas has not been extensively studied due to the lack of observations that combine large-scale mapping and high spatial resolution.

Filaments have long been a focus of research in studies of star formation and the interstellar medium \citep[see recent reviews from][]{Hacar2023, Pineda2023}. They are thought to play a role in transporting gas, collecting it into star-forming structures, and may even reflect global Galactic dynamics \citep{Smith2022,Duarte-Cabral2017}. In the disk of the Milky Way, filaments range from galactic-scale structures spanning lengths of $\sim$100~pc \citep[e.g.][]{Goodman2014, Zucker2018, Ragan2014, Ragan2016} to 0.1~pc scale filaments deep within molecular clouds \citep[e.g.][]{Andre2010, Henshaw2016_IRDC, Wallace2022, Sokolov2017, Barnes2018, Battersby2014}.

Magnetic fields are an important component in the formation and evolution of interstellar filaments. In molecular gas, the orientation and strength of the Plane of Sky (POS) magnetic fields associated with filamentary structures can be studied using polarization from aligned dust grains \citep{Andersson2015,Akshaya&Hoang2024, Sato2024} or Velocity Channel Gradients \updates{of polarization} (VChG; \citealt{Lazarian2018}). The latter has advantages for tracing the magnetic field structure of the CMZ without the contribution from the foreground \citep{2022MNRAS.511..829H, 2022MNRAS.513.3493H}.

A variety of filamentary structures are found to be associated with magnetic fields in the CMZ. Non-thermal radio filaments oriented perpendicular to the Galactic plane have been observed in the CMZ \citep{1984Natur.310..557Y,1987ApJ...322..721Y,2023ApJ...949L..31Y}. The origin of these filaments is debated with several hypotheses and models proposed \citep{2019MNRAS.490L...1Y,2021A&A...646A..66P}. The most prominent Radio Arc filament exhibits an observed length scale around 30~pc \citep{1986ApJ...310..689Y}. Polarized synchrotron emission reveals that the magnetic field orientation in several filaments, e.g., Radio Arc, Snake, Mouse, and N5 \citep{1995ApJ...448..164G,1999ApJ...526..727L,2005AdSpR..35.1129Y,2021ApJ...923...82P} \updab{may be aligned} with the filament spine\updab{s}. The magnetic field strength in these filaments ranges from $\sim$$100~\mu$G to $\sim$$400~\mu$G, as evaluated from Faraday rotation \citep{1995PASJ...47..725I,1995ApJ...448..164G,1987ApJ...322..721Y,2021ApJ...923...82P} and the equipartition method \citep{Yusef-Zadeh2022}.

In this work, we present data from the Atacama Large Millimeter/submillimeter Array (ALMA) CMZ Exploration Survey (ACES)\footnote{\url{https://sites.google.com/view/aces-cmz/}} using the molecular tracer \HNCO. ACES provides a contiguous map of molecular gas across the entire CMZ at high spatial resolution. These images reveal a stunning complexity of filamentary structures throughout the CMZ. We identify two main categories of filamentary structures and quantify the properties of a small sample of structures in each category.
The terminology surrounding filaments is fraught and basic terms remain ill-defined. For this work, we adopt the term `Large-scale Filamentary Structures' (LFs) for elongated structures ($\gtrsim$ 10 pc) that seem to be associated with large-scale orbital gas motions, and we note that these structures may have small gaps, similar to other large-scale filamentary structures in the literature \citep[e.g.][]{Zucker2018, Zucker2015, Schneider2010}. We adopt the term `Small-scale Filamentary Structures' (SFs) for smaller-scale features ($\sim$1~pc) associated with molecular clouds that are typically fully contiguous.
\updates{There are plentiful examples of both types of filamentary structures in the new ACES datset. In this work we neither identify complete samples nor fully define these classes of objects. Rather, our aim is to highlight their existence and prevalence in the CMZ and present a few examples and their physical properties.}

In Sections \ref{sec:data} and \ref{sec:methods}, we describe the data and the methods used in this paper. In Sections \ref{sec:morphology} and \ref{sec:kinematics} we present physical and kinematic properties of the filamentary structures and connect them with the larger orbital streams in PPV space. In Section \ref{sec:magnetic} we provide a detailed comparison with POS magnetic field orientations from the FIREPLACE survey \citep{Butterfield2024, Pare2024}. In Section \ref{sec:chemistry} we investigate the chemical content of the structures by comparison with other molecular species. In Section \ref{sec:discussion} we consider the possible origins for these structures and the uncertainties in our analyses. We end with a summary of our main conclusions in Section \ref{sec:conclusions}, including encouraging further study of these structures, which could provide further understanding of their formation and role in star formation and to help constrain dynamical models of the CMZ.

\section{Data}
\label{sec:data}

\begin{figure*}
\includegraphics[width=1\textwidth]{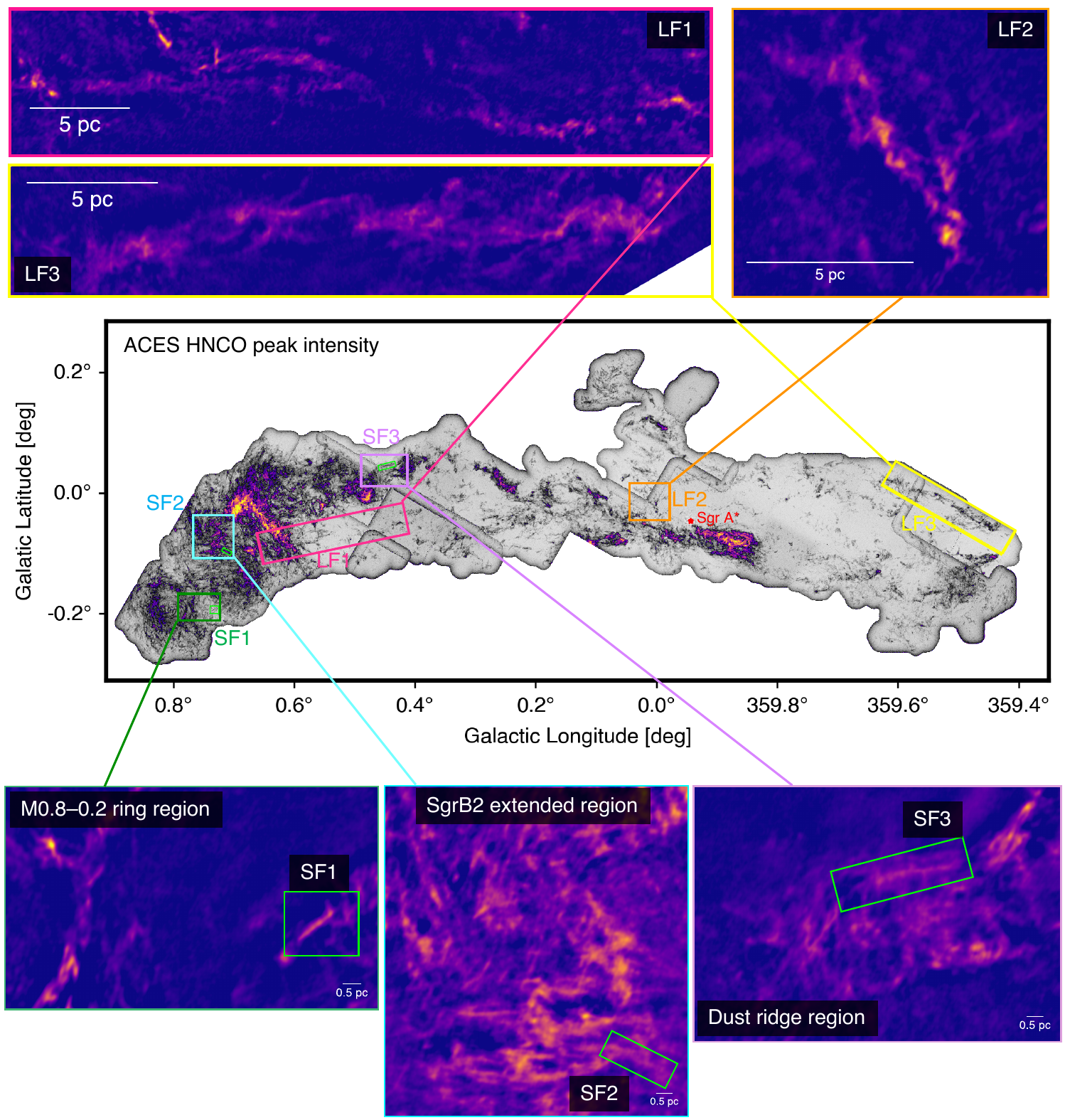}
\caption{The ACES \HNCO~data reveal the ubiquity of molecular filamentary structures in the CMZ on scales from tens to tenths of pc. The central panel shows a peak intensity map of \HNCO~from ACES and our sample selection in the colored boxes. The top three zoom-ins show the three selected Large-scale Filamentary Structures (LFs) in HNCO, integrated over the velocity range of each individual structure (see Table \ref{tab:filament_striation_statistics}). The bottom three zoom-ins show three characteristic regions highlighting the preponderance of Small-scale Filamentary Structures (SFs) in the molecular gas of the CMZ. Within these regions, we select one characteristic SF for quantitative analysis. The zoom-ins on the bottom show HNCO integrated over the velocity range of the selected SFs (see Table \ref{tab:filament_striation_statistics}).}
\label{fig:overview}
\end{figure*}

\subsection{ACES Survey and data reduction}
\label{sec:aces_overview}

The ALMA CMZ Exploration Survey (ACES) is a Cycle 8 Band 3 Large program on ALMA (2021.1.00172.L, PI: S. Longmore). ACES observed the majority of molecular gas in the CMZ spanning approximately $-0.6$\degree$<\ell<0.9$\degree~and $-0.3$\degree$<b<0.2$\degree~with a total area of 1286 square arcminutes in both broad-band continuum and spectral lines at 3mm. The data presented in this work are synthesized images incorporating data from the 12m, 7m, and total power array of ALMA. The spatial resolution (full width at half maximum) of ACES data is about 3\arcsec~(or 0.1~pc at the assumed Galactic Center distance used throughout this work of 8.2~kpc; \citealt{Reid2019,Gravity19,GRAVITYCollaboration2021}).

The ACES data cover six spectral windows: two medium (469 MHz) width from 86 -- 86.5 and 86.7 -- 87.1~GHz, two broad (1.875 GHz) width from 97.66 -- 99.54 and \mbox{99.56 -- 101.44~GHz,} and two narrow (59 MHz) width centered on \HNCO \mbox{($\nu_\mathrm{rest}=87.925238$~GHz)} and \hcop (1--0)  \mbox{($\nu_\mathrm{rest}=89.18852$~GHz)}.
The ALMA pipeline produced measurement sets using the CASA 6.4.1.12 pipeline 2022.2.0.64 version. The data were imaged using CASA 6.4.3-2 with small modifications, as described \mbox{in \citet{Ginsburg2024}.}
Full survey details are presented in ACES Paper I \citep{Longmore2025} providing an overview of the survey. In ACES Paper II, \citet{Ginsburg2025}, we describe the continuum data reduction details and continuum maps. In ACES Paper III, \citep{Walker2025b}, we describe the molecular line data reduction and present the \HNCO~and \hcop (1--0) data spectral lines that were observed with an enhanced spectral resolution. In ACES Papers IV and V, \citet{Lu2025} and \citet{Hsieh2025}, respectively, we present the intermediate-width and broad spectral line observations.

In this work, we primarily focus on the \HNCO~spectral line at a frequency of 87.925237 GHz. 
\updates{In an extensive investigation of 3mm spectral line tracers based on the \citet{Jones2012} MOPRA data, \citet{Henshaw2016a} identified \HNCO~as a promising tracer for dense molecular gas, and used this tracer to build a comprehensive position-position-velocity decomposition of the CMZ that is still used today. Future analysis of the ACES data will focus on the differences between molecular line tracers and hopefully identify the physical structures and conditions indicated by each spectral line. However, at the moment, we chose the same tracer as \citet{Henshaw2016a} as a first look at the position-position-velocity structure of the CMZ at high resolution.}

In the ACES data, the \HNCO~line has a velocity resolution of 0.2~km~s$^{-1}$ (see the additional lines used in this work and their velocity resolutions in Table \ref{tab:chemistry_data}). This line is generally thought to trace the dense, molecular gas in the CMZ and outline its key structural features \citep{Henshaw2016a}. \HNCO~is also a tracer of low-velocity shocks \citep{Martin2008,Kelly2017}, so its detection in the CMZ is associated with the presence of widespread low-velocity shocked gas across the GC \citep{Martin-Pintado1997}.

We compare the features identified in \HNCO~with a number of different molecular lines observed with ACES as listed in Table \ref{tab:chemistry_data}. We provide some references about what these molecular lines may trace in the final column of Table \ref{tab:chemistry_data}, but their precise physical significance is still a subject of active research.

\begin{table*}
\caption{ACES complementary observations. Spectroscopic parameters and critical densities for molecular line species in ACES.}
\label{tab:chemistry_data}
\begin{tabular}{lccccccc}
\hline
\textbf{Species} & \textbf{$\delta$v} & \textbf{Frequency} & \textbf{E$_\mathrm{u}$/k$_\mathrm{B}$} & \textbf{A$_{\rm ul}$} & \textbf{n$_{\mathrm{cr,u}}$(100~K)} & \textbf{Ref. coll. rates}   & \textbf{Tracer} \\
               &  \textbf{(\kms)}   & \textbf{(GHz)}             & \textbf{(K)}                  & \textbf{10$^{-5}\cdot$(s$^{-1}$)} & \textbf{10$^4\cdot$(cm$^{-3}$)} & \textbf{para-H$_2$} & \\
\hline
\HNCO & 0.21    &  87.92523 & 10.6 & 0.9 & 1.9 & a & low-velocity shocks/dense gas$^h$  \\
$^{3}\Sigma$ \SO      &  3.0    &  99.29987 & 9.2 & 1.1 & 3.5 & b & shocks/dense gas$^m$ \\
\HCtrN & 2.9 & 100.07630 & 28.8 & 7.8 & 7.3 & c & low velocity shocks/dense gas$^n$ \\
\SiO     &    1.7     &  86.84696 & 6.3 & 2.9 & 4.3 & d & high-velocity shocks$^k$ \\
\HthCN & 1.7    &  86.33992 & 4.1 & 2.2 & 4.4 & e & dense/moderate-density gas$^j$ \\
\HthCOp  & 1.7    &  86.75430 & 4.2 & 3.9 & 2.2 & f & dense gas/enhanced cosmic rays$^p$\\
\CS      &    3.0     &  97.98095 & 7.1 &  1.7 & 4.3 & g & dense/translucent gas$^i$ \\
\hline
\end{tabular}
\medskip
\par\noindent Notes. 
\mbox{(2) Velocity resolution,} (4) Energy of upper level, (5) Einstein A coefficient, (6) Critical density, $n_{\mathrm{cr,u}}$=$A_{\rm ul}$/$\Sigma_\mathrm{i\neq u} \gamma_\mathrm{ui}$, in collisions with (para-)H$_2$, at~100~K, (8) What each species/line is typically tracing.\\
\textbf{References}, collision rates: (a) \cite{Sahnoun2017}; \mbox{(b) \cite{Lique2007};} (c) \cite{Faure2016}; (d) \cite{Balanca2018}; \mbox{(e) \cite{Hernandez-V2017}; (f) \cite{DenisA2020} (g) \cite{Denis-A2018}.}  \\

\textbf{References}, tracers: (h) \cite{Martin2008,Marcelino2010,Kelly2017}; (i) \cite{Jimenez-Serra2005,Bayet2009}; \mbox{(j) \cite{Tanaka2018};} (k) \cite{Martin-Pintado1992,Kelly2017}; (l) \cite{Pety2017}; \mbox{(m) \cite{Pety2017,Tanaka2018,Santa-Maria2021,Segal2024};} \mbox{(n) \cite{Martin2008,Goldsmith2017,Rod-Baras2021}.}
\end{table*}

\subsection{FIREPLACE data}
\label{sec:other_data}
An important element in characterizing the LFs and SFs is how these structures align with local magnetic fields. We compare the orientations of the targeted LFs and SFs with the magnetic field directions inferred from the recent Far-InfraREd Polarimetric Large Area CMZ Exploration (FIREPLACE) survey that observed the entire CMZ at 214 $\mu$m with SOFIA/HAWC+ at a resolution of 19.6\arcsec  \citep{Butterfield2024,Pare2024}.

We follow the significance cuts employed by \citet{Pare2024} to only consider magnetic field pseudovectors that satisfy the significance conditions: $I_{214}/\sigma_I >$ 200, $p/\sigma_p >$ 3.0, and $p_\% <$ 50 where $I_{214}$ and $\sigma_I$ are the 214 $\mu$m emission and uncertainty, $p$ and $\sigma_p$ are the 214 $\mu$m polarization intensity and uncertainty, and $p_\%$ is the percentage polarization ($p/I\times100)$. These significance cuts align with standard SOFIA/HAWC+ practice \citep{Harper2018}.

FIREPLACE observations provide \updates{a POS component of} the magnetic field \updates{integrated} along the line of sight, weighted by the gas density, and is dominated by the high density structures in the GC \citep{Pare2025}. This is especially true for the LFs and SFs observed in this work, since HNCO traces high density gas.

\section{Methods}
\label{sec:methods}

\begin{figure*}
\centering
\includegraphics[width=0.8\textwidth]{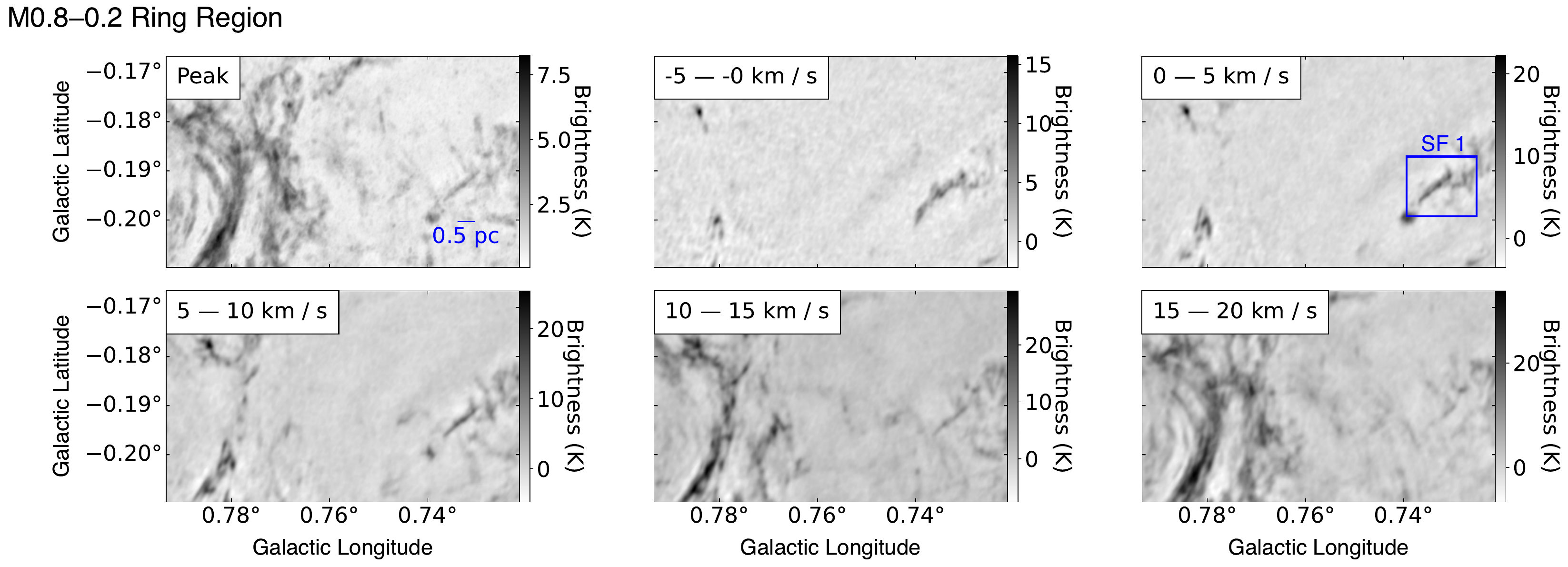} \\
\includegraphics[width=0.8\textwidth]{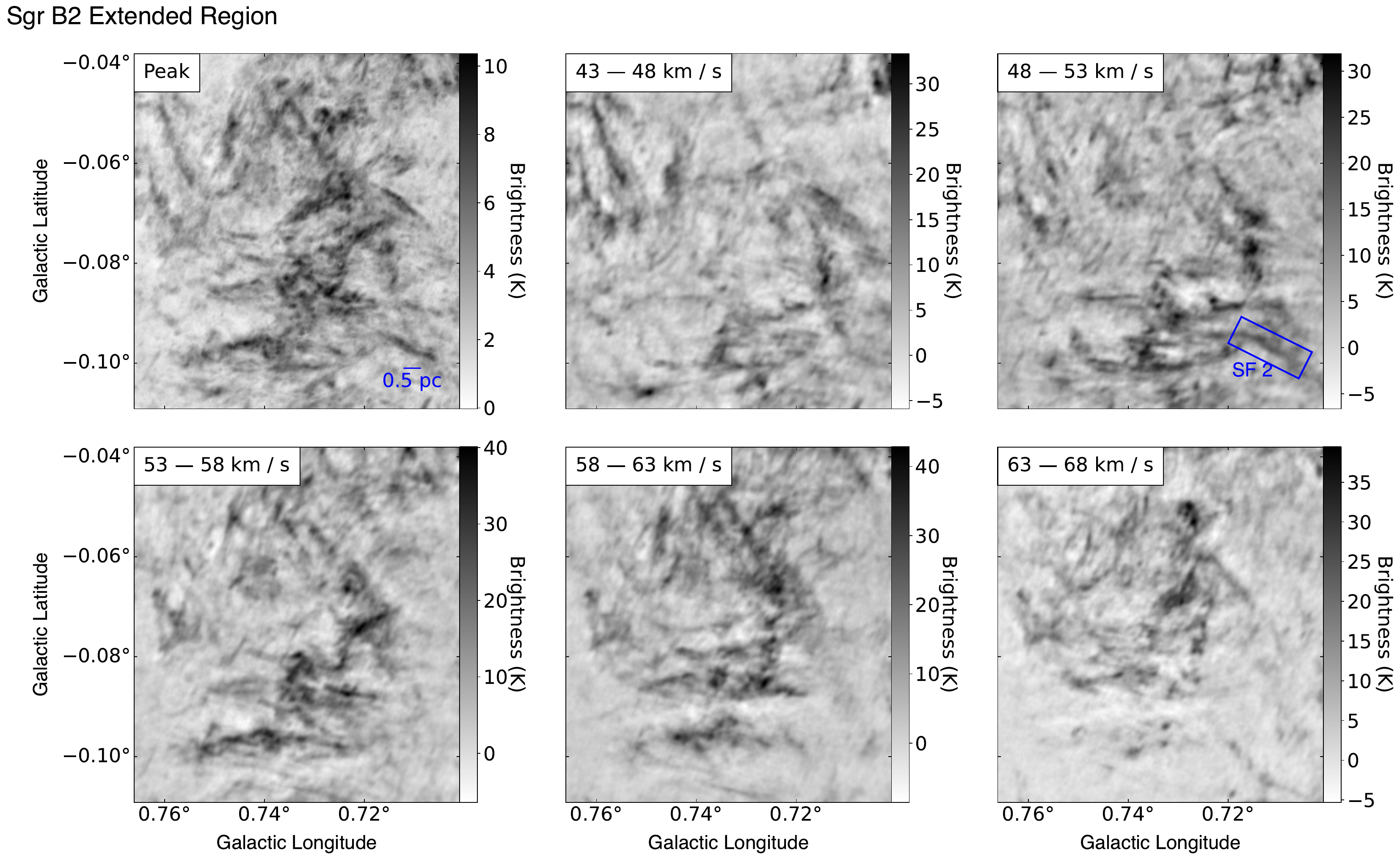} \\
\includegraphics[width=0.8\textwidth]{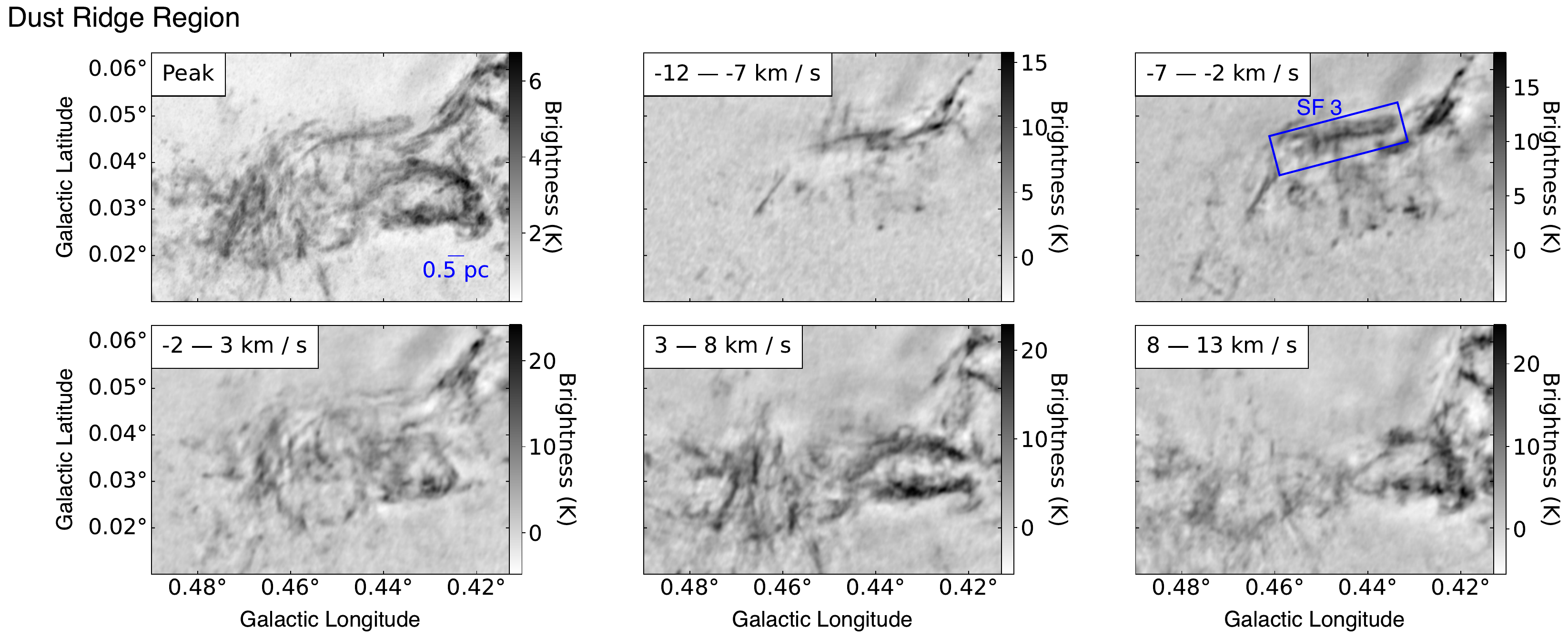}
\caption{Small-scale filamentary structures (SFs) are ubiquitous in molecular gas in the CMZ as traced with \HNCO~by ACES. This image shows a zoom-in on the three cyan SF fields displayed in Figure \ref{fig:overview}, the M0.8–0.2 ring region \citep{Nonhebel2024} (\textit{top}), the SgrB2 extended region (\textit{middle}), and the Dust Ridge region (\textit{bottom}). For each zoom-box, we display the peak intensity over the entire HNCO cube in the top left panel, then step through 5 kms$^{-1}$ intervals with the corresponding moment 0 map of the velocity range displayed in subsequent panels. We highlight the location of SFs 1, 2, and 3 within the map in blue boxes, at the velocity most closely associated with each SF.}
\label{fig:striation_ubiquity}
\end{figure*}

\subsection{Sample Identification and Terminology}

\updates{Small-scale} filamentary structures are \updates{evident and widespread} in the ACES \HNCO~moment 0 and peak intensity maps, highlighted in the zoom-in panels of Figures~\ref{fig:overview} and \ref{fig:striation_ubiquity}. These structures exhibit a wide range of physical sizes and pervade the entire CMZ. We highlight two size scales of filamentary structures found in the ACES \HNCO~data which were most apparent to the authors on first study. The larger-scale ($\gtrsim$10~pc) of these structures, which we refer to as ``Large-scale Filamentary structures'' (hereafter LFs), appear to trace branches of the CMZ orbital streams and may contain small gaps. The smaller-scale ($\sim$1~pc) objects, which we refer to as ``Small-scale Filamentary structures'' (hereafter SFs), appear morphologically similar to the LFs on smaller spatial scales, and often exist near or within molecular clouds. We choose generic terms to avoid confusion with other filament-like structures defined in the literature. It is possible that the chosen sample describes objects of differing morphology and origins. Additionally, our analysis doesn't preclude the possibility of a continuum of filamentary structures over size scales.

\updates{In this work, we do not present a rigorous filamentary structure classification scheme, complete samples of LFs or SFs, nor clear distinctive object class definitions. Our aim is simply to highlight the discovery of the prevalence of LFs and SFs in the CMZ and to present physical and kinematic properties towards a few example structures. To that end,} we identify three LFs and three SFs for quantitative analysis that serve as representative examples of the two populations. \updates{These representative examples were selected through visual inspection by the team to highlight the range of structures observed and are in no way meant to be comprehensive.}

We describe the method to identify these representative LFs and SFs below and compare this with an automated Rolling Hough Transform (RHT) method to identify LFs in a follow-up paper \citep{Pare2025b} at the end of this section. 
The initial inspection of the data to identify these structures was done using a spatially down-sampled version of the ACES \HNCO~data cube due to the prohibitively large size of the full resolution data cube. The lead author led an initial by-eye selection of about 10 structures in each category. Together with members of the paper sprint team, we communally chose three representative objects in each sample for detailed study. After a by-eye determination of the velocity extents for the LFs and SFs, sub-cubes with enhanced spatial and spectral resolution were created for the chosen samples. The full resolution sub-cubes were used for the remainder of the analysis.

The selected LFs and SFs are shown in Figure~\ref{fig:overview}. The main image shows the ACES \HNCO~peak intensity in the background with the zoom-in boxes toward each object in our sample. In the zoom-in boxes we show the HNCO data in each region integrated over the velocity range of the structure (see Table \ref{tab:filament_striation_statistics}). The three LFs are shown in colored boxes spanning the rough size of the full filamentary structures. We highlight the three key regions that demonstrate the ubiquity of SFs in the HNCO data in colored boxes (shown in more detail in Figure \ref{fig:striation_ubiquity}) with the individual SFs selected for quantitative analysis shown in green boxes. \updab{The zoom-in boxes in Figure \ref{fig:striation_ubiquity} highlight regions that seem to contain many dozens of features similar in appearance to the three example SFs selected for analysis. The LFs, on the other hand, are larger, with maybe only a dozen or so similar features apparent in the CMZ \citep[see][for more]{Pare2025b}}. One of the regions is near the M0.8–0.2 ring structure located to the Galactic South of Sgr B2 \citep{Butterfield2024b, Nonhebel2024, Tsuboi2015, Mills2017,Martin2008}, one is the outer extended part of the Sgr B2 region, and one is located within the dust ridge between clouds d and e/f \citep{Immer2012, Battersby2020, Walker2018}. Within each of these regions which each contain many SFs, we chose one SF for quantitative analysis which we label as SFs 1, 2, and 3. SF 1 was particularly chosen as it was slightly isolated from most of the emission in the chosen area. The three regions containing the SFs in Figure~\ref{fig:overview} are further highlighted in Figure \ref{fig:striation_ubiquity}, to show a variety of SFs in different spectral bins. The selected LFs and SFs are located throughout the CMZ and demonstrate a variety of properties. Each of the SFs are observed over multiple velocity channels, and are not due to velocity crowding effects
(see Section~\ref{sec:velocity_crowding}). All of the structures used for our study show evidence for being in the CMZ (see Section~\ref{sec:discussion_inthecmz?}).

In a follow-up paper from the ACES team \citep{Pare2025b}, we use a Rolling Hough Transform (RHT) to systematically identify a larger sample of LFs. As can be seen in Figures 3, 8, and 9 of \citet{Pare2025b}, all three LFs identified in this work are clearly detected with the RHT method, with our LF 1 corresponding to their LF 1b, LF 2 to LF 2a, and LF 3 to LF 3. Since the identification methods are different, the boundaries of the structures vary, but overall have 70\% overlap. The agreement of two independent methods in the identification of the LFs demonstrates their fidelity.

\subsection{Spectral Line Analysis and Velocity Crowding}
\label{sec:velocity_crowding}

When identifying filamentary structures in molecular gas, the width of the velocity channel is a crucial factor. \updates{Theoretical work \citep[such as][]{2000ApJ...537..720L, 2023MNRAS.524.2994H} has shown that when we transform from the known Position–Position–Position (PPP) space to Position–Position–Velocity (PPV) space}, gas at different LOS positions \updates{can have} similar LOS velocities \updates{and therefore} be aggregated into the same PPV space location. This overlap causes the intensity distribution of molecular gas in PPV space to appear more crowded and statistically altered, a significant effect in the turbulent and magnetized ISM. Referred to as velocity caustics, this phenomenon has been extensively discussed by \cite{2000ApJ...537..720L}, \cite{2016MNRAS.461.1227K}, \cite{2021ApJ...910..161Y}, and \cite{2023MNRAS.524.2994H}. It can lead to the formation of non-real SFs parallel to magnetic fields in narrow velocity channels, particularly when the channel width is smaller than the turbulent velocity dispersion.

To ensure proper identification of SFs that arise from real density structures in spectral line cubes, increasing the width of the channel map in excess of the turbulent velocity dispersion is recommended. The velocity crowding is no longer dominant if the channel width exceeds the turbulent velocity dispersion.
To avoid velocity crowding in our analysis, we use integrated moment 0 maps and velocity maps, constructed from a minimum of 5~\kms to a maximum corresponding to the upper limit of the spectral axis for the \HNCO~cube (104~\kms), to avoid per-channel morphological analysis in the identification of our structures.

\begin{figure*}
\includegraphics[width=1\textwidth, trim=20mm 5mm 20mm 0, clip]{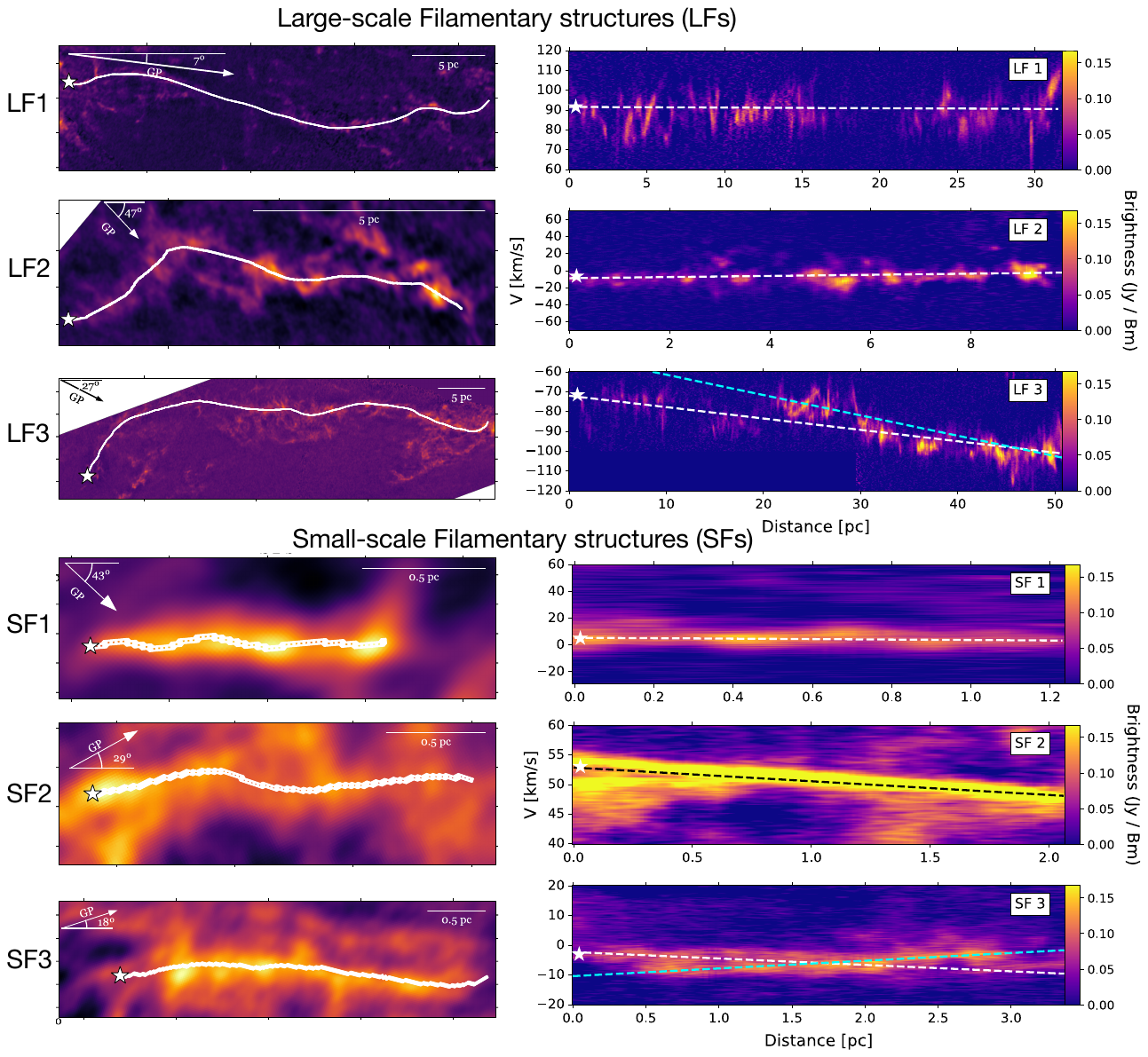}
\caption{A morphological comparison of the three LFs and SFs is presented in the \textit{left} panels, which highlights both their similarities (extended filamentary structures coherent in PPV space) and their differences (size scales). The \textit{right} panels show position-velocity diagrams extracted along the spines (shown as the white line along each filament in left panels, described in Section \ref{sec:method_morphology}), and show largely coherent structures in PV space. The \textit{left} panels show \HNCO~moment 0 maps (over the velocity ranges listed in Table \ref{tab:filament_striation_statistics}) of each structure rotated so that they are all roughly parallel. The position angle of the filament spines with respect to the Galactic plane is given in the top left of each image. The white stars denote the starting position used for the PV diagrams shown in the \textit{right} panels. The white (black for SF 2) dashed line on the PV plots show the fit used to calculate the filament velocity gradients (Table \ref{tab:filament_striation_statistics}). The blue dotted lines in LF3 and SF3 represent alternate slopes (1.0 and 2.6 $km s^{-1} pc^{-1}$ respectively) to the velocity gradients provided in Table \ref{tab:filament_striation_statistics}. The SFs show clear contiguity over their short lengths, while the LFs have small gaps in the velocity contiguity over their 10+ pc lengths (most notably LF 1 in the center).  }
\label{fig:statistics_pv}
\end{figure*}

\subsection{Measuring Structure Morphology}
\label{sec:method_morphology}
We measured morphological properties for the selected sample of LFs and SFs using the python packages \texttt{\texttt{FilFinder}} \citep{Koch2015} and \texttt{\texttt{RadFil}} \citep{Zucker2018_2}. Both packages require image masks to isolate the filament from the background. To produce these masks, we used the multi-dimensional data exploration and visualization tool \texttt{glue} to create hand-drawn masks on the integrated velocity subcubes of the \HNCO~map around our filamentary structures. Hand-drawn masks were preferred over e.g. a sigma cut, since the statistical properties of noise in the ACES data were still being established at the time. Emission outside the masks is removed for the quantitative analyses.

We use \texttt{FilFinder} to define the structure's spines on the \HNCO~moment 0 maps as shown in the left panel of Figure \ref{fig:statistics_pv}. These spines define the location and path of the filament and are created by reducing structures to a ``skeleton" using a medial axis transform based on average intensity around the density ridge of the filament \citep{Koch2015}. When defining the masks, the filaments diverged in some places. For example, LF 2 appears to have two potential paths for the spine to follow based on extended nature of the structure (see Figure \ref{fig:statistics_pv}). When creating our masks, we determined that the lower path of the filament was preferred as it traces the connected structure of the filament in ppv space. Table \ref{tab:filament_striation_statistics} reports on the physical properties of the structures. The lengths in this Table were calculated from the lengths of the spines at the assumed 8.2~kpc distance to the Galactic Center \citealt{Reid2019,Gravity19,GRAVITYCollaboration2021}.

We computed the Full Width at Half Maximum (FWHM) widths of the filamentary structures perpendicular to the defined spines using \texttt{RadFil} \citep{Zucker2018_2}. The widths are computed using the masked \HNCO~moment 0 maps. \texttt{RadFil} compiles the perpendicular profiles of all points along the spine, centers them on the maximum value and performs a Gaussian fit to the average of the profiles in the radial direction. Taking the FWHM of this Gaussian fitting, \texttt{RadFil} deconvolves the width using the beamwidth, which we defined to be 3.19''. This deconvolved FWHM is related to the non-deconvolved FWHM as $\rm FWHM_\mathrm{deconv}=\sqrt{\mathrm{FWHM^{2}-B^{2}}}$, where B is the beamwidth. The deconvolved FWHM values are presented in Table \ref{tab:filament_striation_statistics}. For both LFs and SFs, uncertainties in the FWHM are calculated based on standard deviations of the widths and equal $\sim 0.07$pc, though we note this approach could hide variations in radial profiles \citep[see e.g.,][] {Schmiedeke2021}. The position angle to the Galactic Plane was calculated relative to constant latitude (0\textdegree). The filament FWHMs and position angles are reported in Table \ref{tab:filament_striation_statistics}.

We also determined an axis ratio of each of the filaments. To do this, we first compute the corresponding geometric width of each filament, since the FWHMs reported are fractions of a Gaussian structure and are not directly comparable to geometric lengths. We define an effective radius, $R_{\rm eff}$, by equating the area under a Gaussian curve with a normalization of one ($2\pi\sigma^{2}$) to the area of a circle ($\pi$$r^{2}$). Using the standard deviation defined by a Gaussian curve, we then find the effective radius to be:
\begin{equation}
\label{eq:reff}
  R_{\rm eff}=\frac{\rm FWHM}{2\sqrt{\ln 2}}
\end{equation}
where ${\rm FWHM} = 2 \sqrt{2\ln 2} \, \sigma$.
Our geometric filament diameters are therefore 2$R_{\rm eff}$, which is used to calculate the axis ratios (geometric length / diameter) reported in Table \ref{tab:filament_striation_statistics}.

We report the median column density, N(H$_2$), of each structure in Table \ref{tab:filament_striation_statistics}. They were calculated using the Herschel-based column density maps from \citet{Battersby2025a}, by using our structure masks (re-gridded to match the pixel scaling of the column density map) for each structure and computing a statistical median value.

Finally, we measure the curvature and linear nature of the SFs and LFs. To compute these values, we used \texttt{FilFinder}'s curvature calculator using the default parameters (radius of circular region is 10 pixels, number of bins in which $\theta$ is divided is 180, and background subtractions of the 25th percentile). \texttt{FilFinder} calculates the curvature and orientation of the filamentary structures by implementing the Rolling Hough Transform of the spine and the branches. These values are reported in Table \ref{tab:filament_striation_statistics}. For the curvature of the spine as a whole, it computes the interquartile region about the mean. More information about the Rolling Hough Transform and curvature can be found in \citet{Clark2014} and \citet{Koch2015}.

\subsection{Kinematic Structure}
\label{sec:method_kinematic}
We study the kinematic structure of the LFs and SFs. For both types of objects, we map them in 2-D position-velocity (PV) space as well as 3-D PPV space. We aim to uncover how contiguous they are in velocity space as well as to understand their position within the larger known CMZ orbital features and molecular clouds.

High-resolution ACES \HNCO~sub-cubes were created for each of the six identified structures, covering the velocity channels in which a given structure was seen in the original \HNCO~maps (see velocity ranges used in Table \ref{tab:filament_striation_statistics}). We generated PV diagrams using these high-resolution sub-cubes (Figure \ref{fig:statistics_pv}).
We used the spines identified using \texttt{FilFinder} in Section \ref{sec:method_morphology} and extracted a PV slice along the spine using the \texttt{PVextractor} python package \citep{Ginsburg2016pvextractor}. We averaged over the spine width, 20 arcseconds for LFs and 10 arcseconds for SFs, to create the PV diagrams. We then find the approximate slope of the bright PV emission by drawing a best fitting line between the start and end points of each PV structure to determine a velocity gradient, shown as the dashed line in the right panels of Figure \ref{fig:statistics_pv}. As seen in Figure \ref{fig:statistics_pv} the LFs appear largely contiguous in PV space, with some gaps over their extended lengths. The SFs are fully contiguous, including SF 2 which is remarkably continuous in PV space, despite its more diffuse presentation in the moment 0 map. \updates{Though SF 2 shows a small divergence in the moment 0 map, the structure is contiguous and bright across the entire spine in PV space.} In Table \ref{tab:filament_striation_statistics} we report on the approximate velocity gradients in each of the structures. We note possible different slopes for portions of LF 3 and SF 3 (1.0 and 2.6 $km s^{-1} pc^{-1}$ respectively) and show these with blue dotted lines in Figure \ref{fig:statistics_pv}.

We calculate the \HNCO~central velocities and line widths (FWHMs). The line width maps were obtained by taking the square root of the moment 2 (velocity dispersion) maps of the \HNCO~sub-cubes. However, given the large velocity ranges in the LFs and therefore multi-peaked spectra, it was necessary to apply both spectral and spatial masks to isolate the emission that is both associated with the structure and above the noise level in the sub-cube. To create the spectral mask, we first visually identified a flat part of the spectra where there was no emission, in order to approximate an average noise value in the spectral dimension, $\sigma$. A 3-D noise mask is then constructed, which cuts out any pixels below a value of $3\sigma$ across the entire sub-cube. We then employ the \texttt{scipy.ndimage} library to first remove the jagged edges of the mask via the \texttt{binary\_erosion} function, and then expand the mask to catch any signal that may have been cut in the previous step using \texttt{binary\_dilation}. The result is a spectral mask with smoothed edges that is more likely free from sharp artifacts from just a simple $3\sigma$ cut.
Once the sub-cube had been masked in the spectral dimension, we calculated the line width FWHM using the \texttt{spectral\_cube} python package \citep{Ginsburg2019spectralcube}, which finds the line width by taking the square root of the moment 2 map (M$_{2}$) such that the FWHM is defined as:

\begin{equation}
  FWHM = \sqrt{M_{2}}\sqrt{8\mathrm{ln}2}
\end{equation}

Lastly, we apply a spatial mask to the calculated FWHM maps to isolate the shape of the structures in the \HNCO~moment 0 maps. We extract the spatial components of the mask from the integrated intensity (moment 0) maps. We apply a Gaussian smoothing of 1 beam for all structures, and mask the emission below a 4$\sigma$ cut (where $\sigma$ is the local noise value). The resulting masks cover the brightest parts of the moment 0 emission, but still include superfluous, bright emission unassociated with the structures. The extraneous emission is eliminated by bitwise multiplication of the smoothed contour-derived masks with the hand-drawn masks created using \texttt{Glue} as described in Section \ref{sec:method_morphology}, leaving us with a moment 0 mask only associated with bright emission from each structure. The spatial masks are applied to the FWHM maps to isolate the regions associated with each LF or SF. 
The average spectral line FWHMs for each structure are reported in Table \ref{tab:filament_striation_statistics}. The spectral line FWHMs fall in the range of 4-9~km/s, which lie within the standard size-linewidth relationships for the CMZ (see Section~\ref{sec:discussion_inthecmz?}). We note that \citet{Pare2025b} find larger line widths for the LFs studied here \updates{(5-8$\times$ larger)}. \citet{Pare2025b} computed simpler, \updates{moment 2 based} structure estimates on larger-scale convolved data, resulting in larger line widths.

To analyze the association of our LFs and SFs with the overall orbital motions of CMZ gas, we compare the structures' spatial extents and their average weighted velocities by plotting them in PPV space in Figure \ref{fig:ppv}. We refer the reader to \citet{Walker2025} and related resources for an interactive 3D CMZ viewer. We apply the combined spatial masks detailed above to the moment 1 maps to extract centroid velocities at each pixel. We compare our extracted ($\ell$, $b$, v) points for each pixel in the \HNCO~sub-cube with the \texttt{scousepy} spectral decomposition of Mopra \HNCO~data from \citet{Henshaw2016a}, discussed further in Section \ref{sec:kinematics}.

\subsection{Measures of the Magnetic Field Alignment}
\label{sec:method_magnetic}
The alignment between the FIREPLACE magnetic field and the \HNCO~filaments is first assessed qualitatively by-eye as seen in Figures \ref{fig:Bfield_fil} and \ref{fig:Bfield_str}. It is then assessed quantitatively using the Projected Rayleigh Statistic (PRS) method \citep{Jow2018,Soler2019}, briefly summarized here. In this paper we apply the PRS on sets of relative angles between the magnetic field orientation derived from FIREPLACE and the intensity structures of \HNCO~emission for the targeted filamentary structures. The intensity structures are characterized by their moment 0 gradient, which is perpendicular to the iso-intensity contours. We define the relative angle following \cite{Soler2017}:
\begin{equation}
  \phi =\tan^{-1}\left(\frac{|\nabla I\times\hat{E}|}{\nabla I\cdot\hat{E}}\right), \label{eq:phi}
\end{equation}
where $\nabla I$ is the gradient of the moment 0 distribution and $\hat{E}$ is the unit polarization pseudo-vector, which is perpendicular to the magnetic field.

We use the normalized alignment measure (AM) parameter introduced in \cite{Gonzalez-Casanova2017} and \cite{Lazarian2018}:
\begin{equation}
  AM=\left\langle\cos 2\phi \right\rangle
\end{equation}
where AM is the alignment measure evaluated over a set of $n$ relative angles ($\phi$) as defined in Equation \ref{eq:phi}. $n$ is chosen to be all the relative angles coincident with the filament's emission. A value of $\mathrm{AM}>0$ indicates a preferentially parallel alignment of the magnetic field with the intensity structure. A value of $\mathrm{AM}<0$  indicates a preferentially perpendicular alignment of the magnetic field with the intensity structure.

\begin{table*}
\caption{Small- and Large-Scale Filamentary Structure Properties}
\label{tab:filament_striation_statistics}
\setlength{\tabcolsep}{4pt}
\renewcommand{\arraystretch}{1.2}
\begin{tabular}{cccccccccccccc}
\hline
\textbf{ID} & \multicolumn{3}{c}{\textbf{Centroid}} & \parbox{1.2cm}{\centering \textbf{Velocity \\ Extent}} & \parbox{1.5cm}{\centering \textbf{Velocity \\ Gradient}} & \parbox{1.2cm}{\centering \textbf{Line Width \\ (FWHM)}} & \parbox{1.2cm}{\centering\textbf{Median \\ N(H$_2$)}} & \textbf{Length} & \textbf{Width} & \textbf{Curvature} & \parbox{0.8cm}{\centering \textbf{Axis \\ Ratio}} & \parbox{1.2cm}{\centering \textbf{Position \\ Angle}} & \parbox{1.5cm}{\centering \textbf{Magnetic \\ Angle}} \\
\cline{2-4}
& \textbf{($l$, \textdegree)} & \textbf{($b$, \textdegree)} & \textbf{($v$, \kms)} & \textbf{(\kms)} & \textbf{(\kms pc$^{-1}$)} & \textbf{(\kms)} & \textbf{($10^{22}$ cm$^{-2}$)} & \textbf{(pc)} & \textbf{(pc)} & \textbf{(\textdegree)} & & \textbf{(\textdegree)} & \\
\hline
SF 1 & 0.73 & -0.19 & 4.4 & (-5,13) & 1.7 &  7.3 & 5.1 & 1.3 & 0.08 & 26 & 14 & 43 & N/A \\
SF 2 & 0.71 & -0.10 & 51.5 & (46,56) & 2.3 & 4.4 & 13 & 2.1 & 0.12 & 27 & 15 & 29 & 30\textdegree \\
SF 3\textsuperscript{a} & 0.44 & 0.05 & -4.0 & (-11,5) & 0.6 & 6.9 & 4.1 & 3.4 & 0.14 & 26 & 21 & 18 & 90\textdegree\ $\perp$ \\
LF 1 & 0.54 & -0.07 & +99.8 & (50,120) & 0.03 & 4.7 & 3.9 & 32 & 0.16 & 29 & 166 & 7 & 90\textdegree\ $\perp$ \\
LF 2 & 0.01 & 0.01 & -9.6 & (-22,20) & 0.7 & 8.5 & 1.3 & 9.8 & 0.18 & 41 & 46 & 47 & 0\textdegree\ $\parallel$ \\
LF 3\textsuperscript{b} & -0.50 & -0.02 & -96.5 & (-120,-50) & 0.6 & 4.3 & 4.3 & 49 & 0.69 & 34 & 590 & 27 & Mixed\\
\hline
\end{tabular}
\medskip
\par\noindent \textsuperscript{a} SF 3 is associated with the dust ridge.
\par\noindent \textsuperscript{b} LF 3 is associated with the ``wiggles'' region \citep{Henshaw2016b}.
\par\noindent Notes. An overview of filamentary structure properties as defined in Section \ref{sec:method_morphology}. The position angle indicated above is the angle of the filament's spine with respect to the Galactic plane.
\end{table*}

\section{Results}
\label{sec:results}

In this section, we report our results from the analyses described above. In Section~\ref{sec:morphology}, we report on the morphology of the structures and in Section~\ref{sec:kinematics}, we discuss the coherence of the structures in both PV (\ref{sec:results_PV}) and PPV space (\ref{sec:results_PPVspace}). The results of the magnetic field analysis are detailed in Section~\ref{sec:magnetic}, where we discuss the orientation of the LFs with the magnetic field (\ref{sec:results_Bfield_vectors}) as well as the Alignment Measurement and HRO analysis (\ref{sec:results_bfield_histogram}). Lastly, in Section~\ref{sec:chemistry} we report the comparison of the \HNCO~line intensities to other molecular lines found with ACES.

\subsection{The Morphology of LFs and SFs}
\label{sec:morphology}
We report on the physical sizes of the LFs and SFs in Table~\ref{tab:filament_striation_statistics}. By our selection and definition, the SFs have shorter projected lengths, typically a few pc (1.3 to 3.4 pc), while the LFs have lengths of tens of pc (10 to 49 pc). We note that we do not account for projection effects, so these lengths should be considered to be lower limits (more in Section \ref{sec:morphology_uncertainties}). The widths on the other hand are more similar between the structure classes. All filaments range between 0.1 and 0.2 pc in their width (FWHM), except LF 3, with a width of about 0.7 pc. 

The axis ratios are smaller for the SFs (between 14 to 21) than the LFs (between 46 to 590). The curvature for all the SFs are roughly the same (26-27\degree) while the LFs vary from 29 and 41\degree. The position angles reported in Table \ref{tab:filament_striation_statistics} indicate that all the filaments tend to be relatively aligned with the Galactic Plane, with only two structures with position angles greater than 30\degree~(SF 1 at 43\degree~and LF 2 at 47\degree). 

We note that we report different lengths and widths for the LFs in this work than were found in \citet{Pare2025b}. \citet{Pare2025b} provided simpler structure estimates that were computed on convolved data, resulting in different properties. For example, they report larger filament widths than those reported here due to the convolution.

\subsection{The Kinematic Properties of LFs and SFs}
\label{sec:kinematics}

\subsubsection{LFs and SFs are largely coherent in both PP and PV space }\label{sec:results_PV}

We report the central velocities, approximate velocity extents, and velocity gradients for each structure in Table \ref{tab:filament_striation_statistics}. The central velocity is taken from the channel at the peak of emission of within the structure. The velocity extent is the range from the minimum to maximum velocity channel where emission associated with the structure is seen. We show the PV diagrams along the LF and SF spines in Figure \ref{fig:statistics_pv}). The LFs are extended in velocity space, spanning a velocity range of 40 to 70\kms~total. However, since they are also quite long, their velocity gradients are small, ranging from 0.03 to 0.7 \kms pc$^{-1}$. The SFs are much more compact, and only extend from 10 to 18 \kms~in velocity. With their small lengths, though, their velocity gradients range from 0.6 \kms pc$^{-1}$ to 2.3 \kms pc$^{-1}$. Nearby parsec-scale filaments in the Galactic disk have typical longitudinal velocity gradients of $\sim$ 1--2 \kms pc$^{-1}$ \citep{Bally1987, Lu2018, Hu2021}, while the large-scale filaments in the disk show similar or smaller velocity gradients \citep{Ragan2014}. Overall, the velocity extents and gradients of the LFs and SFs seem to be within the typical range expected for the respective scales within the Galactic disk.

The LFs appear more diffuse in both PP and PV space compared to the compact PV structure of the SFs seen in Figure \ref{fig:statistics_pv}; however, they are coherent across their lengths. The LFs show consistent emission across their lengths in the \HNCO~PP and PV diagrams, with small gaps, most notably LF 1 between 16 -- 21 pc ($0.5$\degree\ $\lesssim \ell \lesssim 0.55$\degree). Additionally, LF 3 has brighter emission on the right side of the PV diagram that could be consistent with an alternate slope (dashed cyan line: 1.0 \kms pc$^{-1}$) compared with the main velocity gradient (white dashed line: 0.6 \kms pc$^{-1}$).

The SFs are consistently bright and compact across their lengths in both the PP and PV diagrams (Figure \ref{fig:statistics_pv}). SF 1 has multiple bright peaks throughout. SF 2 has two potential paths in PP space, of which the lower was chosen, which shows bright, coherent emission in PV space. \updates{Such filament twisting, splitting, or merging has been seen in a number of filaments across spatial scales, for example, $>$ 10 pc scale Filament 5 from \citet{Zucker2018} and the $\sim$ 1 pc scale G32.02$+$0.06 filament toward a high-mass star-forming region in \citet{Battersby2014}.} SF 3 also has multiple bright peaks and potentially a slightly diverging structure in PV space, such that two velocity gradients were fit. The white line is reported in Table \ref{tab:filament_striation_statistics} (0.6 \kms pc$^{-1}$) while an alternate, steeper slope can be seen in cyan (2.6 \kms pc$^{-1}$).

\subsubsection{The LFs trace the orbital streams in PPV space}\label{sec:results_PPVspace}

\begin{figure*}
\includegraphics[width=1\textwidth]{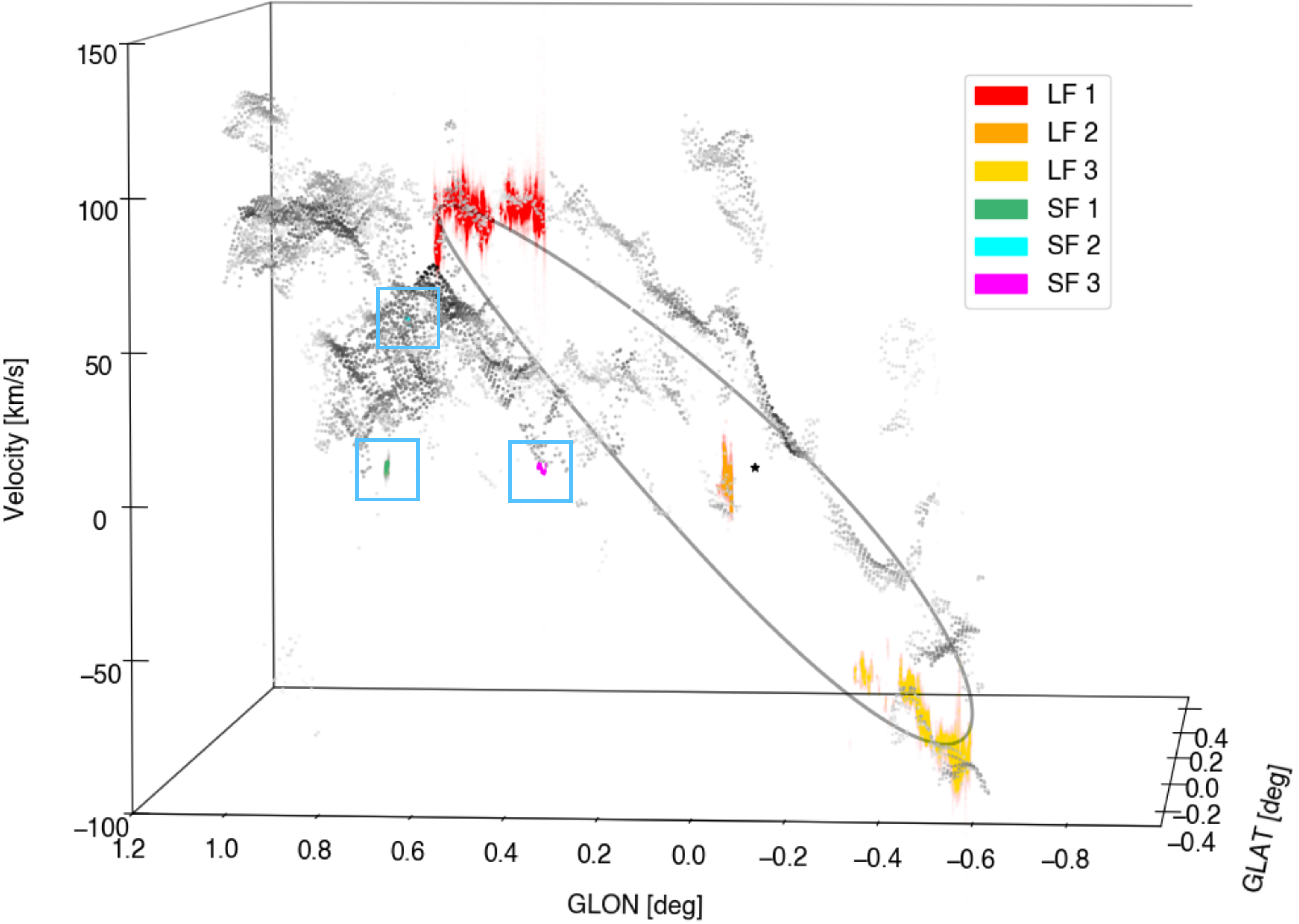}
\caption{The LFs identified in this work appear to trace the CMZ orbital streams, while the SFs are embedded within CMZ cloud structures. This image shows the LFs and SFs identified in this work on top of the kinematic decomposition of Mopra HNCO data created by \citet{Henshaw2016a} (gray background points). The darkness of the gray points is proportional to the peak intensity of each point's spectral component. The closed ellipse orbital model from \citet{Walker2025} is shown as a gray line.  The colored points correspond to LF 1 (red), LF 2 (orange), and LF 3 (yellow), SF 1 (green), SF 2 (cyan), and SF 3 (magenta). Light blue boxes are included to help identify the location of the SFs. The extraction of PPV points for the LFs and SFs is discussed in Section \ref{sec:method_kinematic}. We refer the reader to \citet{Walker2025} and related resources for an interactive 3D CMZ viewer.}
\label{fig:ppv}
\end{figure*}

We compare the large-scale filamentary structures (LFs) in \HNCO~to the orbital streams identified via spectral line decomposition in \citet{Henshaw2016a} (Figure \ref{fig:ppv}) of Mopra \HNCO~data. Similar to other literature LFs \citep[e.g.][]{Zucker2015, Zucker2018, Schneider2010}, these structures show small gaps. The \citet{Henshaw2016a} \texttt{scousepy} decomposition shows the overall view of the dense CMZ gas. In Figure \ref{fig:ppv} the centroids of the individual spectral line components are shown as gray points, with a darkness directly proportional to their peak intensity.
We describe the method for extracting PPV points of the LFs and SFs in Section \ref{sec:method_kinematic}. These LF and SF points are shown in a rainbow of colors, with cyan boxes surrounding the SFs. All of the LFs and SFs are associated with structures in the \citet{Henshaw2016a} decomposition, with the LFs largely on top of the orbital streams. The \citet{Walker2025} CMZ orbital model is shown as a gray closed ellipse in Figure \ref{fig:ppv} for reference.

LF 1 is shown as red points in Figure \ref{fig:ppv} and lies along a dense, high velocity stream that is below the dust ridge in PP space (see Figure \ref{fig:overview}), but closely follows the orbital ellipse from \citet{Walker2025}. The dust ridge is seen in the PPV plot in lower velocity channels with a higher Mopra HNCO \texttt{scousepy} centroid point density. SF 3 (magenta points in Figure \ref{fig:ppv}) is embedded within the dust ridge between clouds d and e/f in PP space, however its central velocity is slightly lower (-4.0 \kms) than cloud d (19 \kms) or cloud e/f (28 \kms) \citep{Battersby2025a, Walker2025}.

SF 2 (cyan points in Figure \ref{fig:ppv}) lies within a complex network of SFs (see middle panel of Figure \ref{fig:striation_ubiquity}) in the Sgr B2 Extended region. This region contains a very high density of points from \citet{Henshaw2016a} that is extended in PPV space. The existence of a small, highly coherent filamentary structure (SF 2) within the immense kinematic complexity of this region, just southeast of Sgr~B2, is worthy of further study.

SF 1 (green points in Figure \ref{fig:ppv}) was selected to be a highly isolated structure just west of the M0.8-0.2 Ring \citep{Nonhebel2024, Butterfield2024b, Tsuboi2015}. In the PPV plot, we confirm that it is isolated and associated with a small cluster of points identified by \citet{Henshaw2016a} in their {\sc scousepy} decomposition. At a central velocity of 4.4 \kms, SF 1 is at a lower velocity than the M0.8-0.2 Ring's systemic velocity of 37.5 \kms~\citep{Nonhebel2024}, Sgr B2 (62 \kms), and Sgr B2 extended (28 or 58 \kms) \citep{Walker2025}. However, its line width (see Table \ref{tab:filament_striation_statistics} of 7.3~\kms) is consistent with a position within the CMZ (see Section \ref{sec:discussion_inthecmz?}).

LF 2 (orange points in Figure \ref{fig:ppv}) shows an intriguing velocity signature: it is relatively close in PPV space to SgrA* and has an extended velocity structure spanning $-$22 to 20 \kms. The velocity extent of LF 2 nearly connects it from SgrA* (lower velocity channels) to the x2 orbital ellipse model from \citet{Walker2025} (upper velocity channels). Such extended velocity features (EVFs) have been hypothesized to be critical exchange points between CMZ orbital families \citep{Sormani2019b}. EVFs in ACES data are the subject of a follow-up publication (Lipman et al., in prep.).  LF 2 is co-located with a small cluster of points identified previously in \HNCO~with Mopra \citep{Henshaw2016a}.

Lastly, LF 3 (yellow points in Figure \ref{fig:ppv}) is at very negative velocities towards the edge of the orbital streams \citep{Walker2025} and follows the kinematically identified structures from \citet{Henshaw2016a} very well. LF 3 is associated with a region of the CMZ, colloquially referred to as the ``wiggles'', identified in \citet{Henshaw2016b} as an area where gravitational instabilities could result in a sinusoidal pattern in velocity space, as can be seen in the PV diagram in Figure~\ref{fig:statistics_pv}.

Overall, LFs 1 and 3 trace our best guess at the current 3D CMZ orbital model well in PPV space, and not just individual points, but over their entire $> 10$ pc lengths. LF 2 also lies near the orbital model, however, its velocity extent may connect the orbital stream with the region surrounding Sgr A*. SFs 2 and 3 are associated with CMZ molecular clouds and SF 1 is isolated in both position and velocity space.

\subsection{Connection with Magnetic Fields}
\label{sec:magnetic}

\subsubsection{Qualitative comparison of POS magnetic field with filament orientations}\label{sec:results_Bfield_vectors}

\begin{figure*}
\centering
\includegraphics[width=0.8\textwidth]{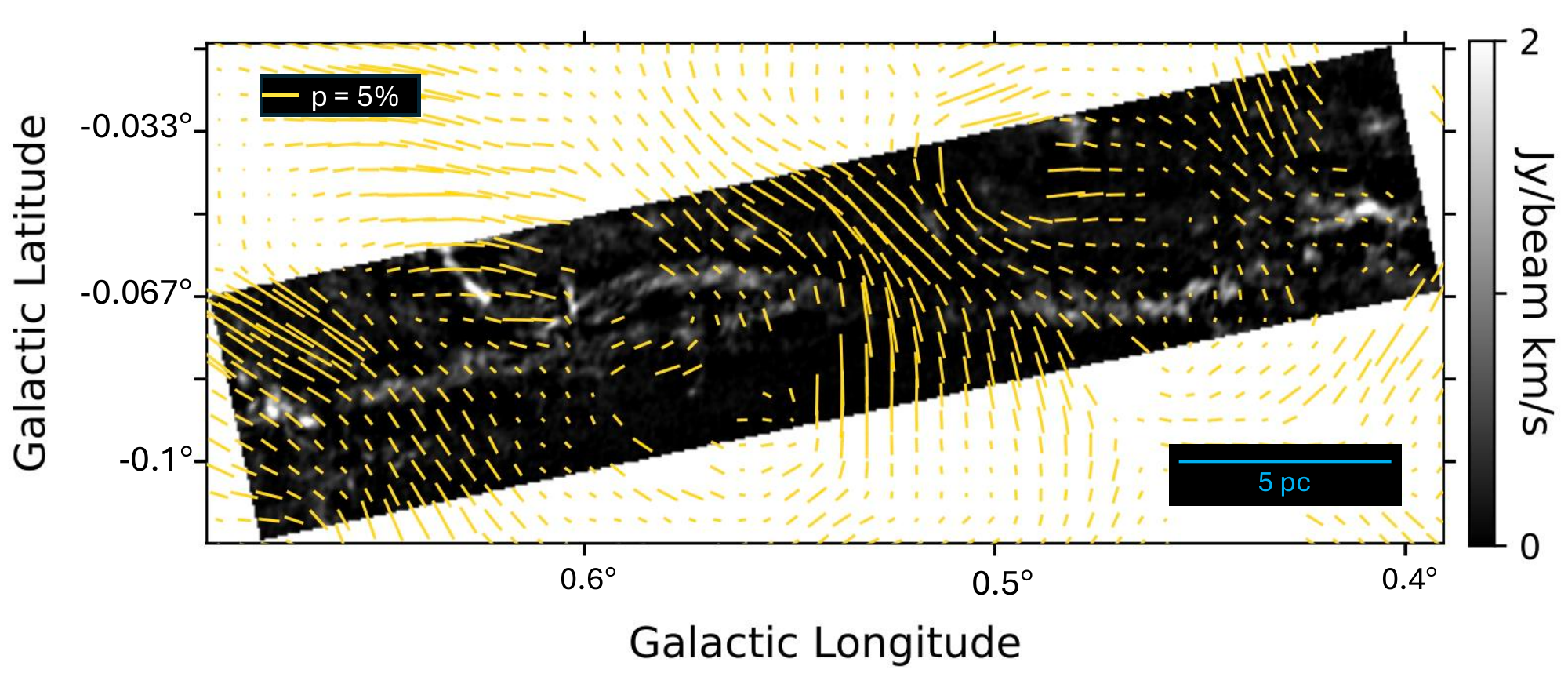}\\
\includegraphics[width=0.5\textwidth]{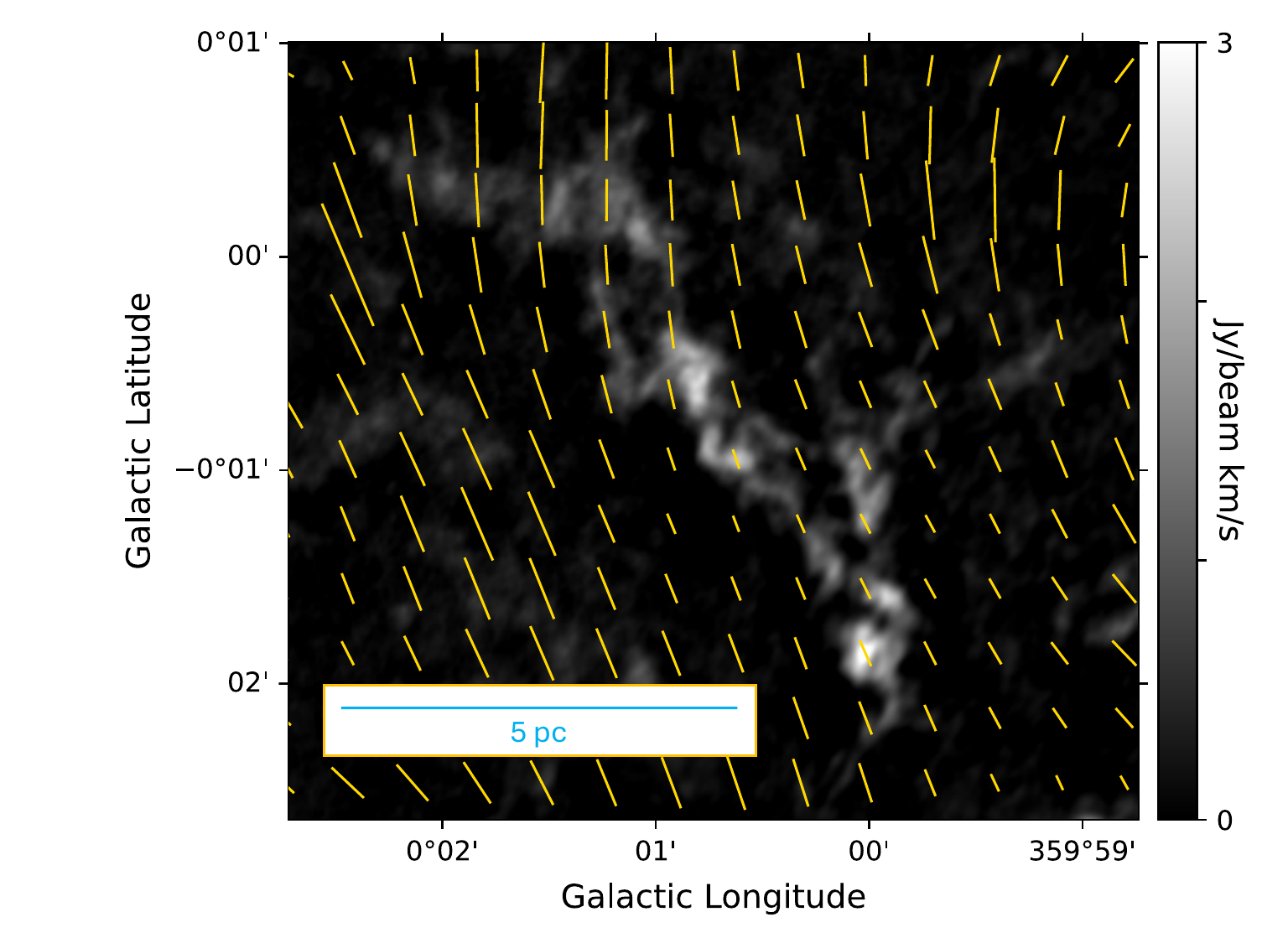}\\
\includegraphics[width=0.8\textwidth]{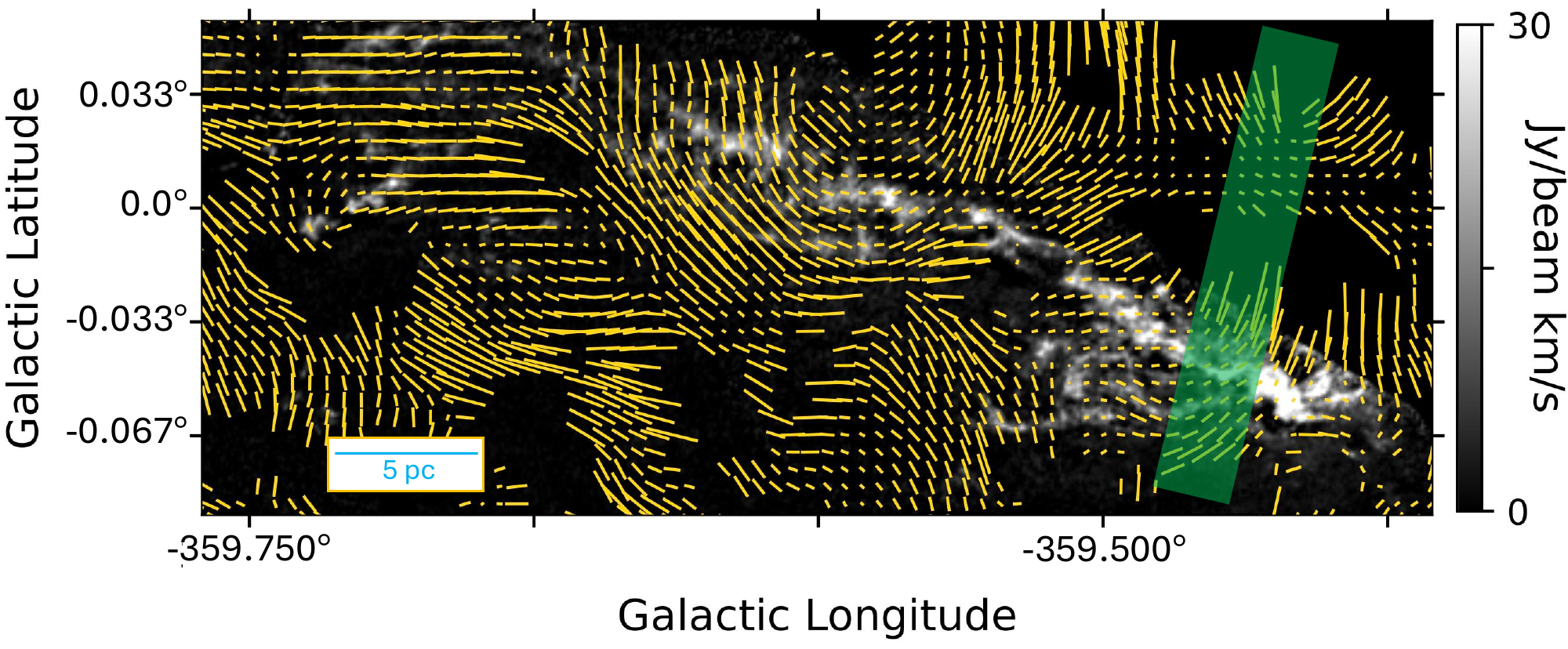}
\caption{A qualitative comparison of POS magnetic field orientations from FIREPLACE (yellow lines) with the LFs shows that LF 1 (\textit{top}) is mostly perpendicular, LF 2 (\textit{middle}) is mostly parallel, and LF 3 (\textit{bottom}) is mixed. Grayscale shows the \HNCO~moment 0 map (integrated over the velocity extent of the structure) with yellow line segments indicating magnetic field pseudovectors derived from the FIREPLACE survey. The line segment lengths are scaled by percentage polarization. \textbf{Inset rectangles indicate the length of a pseudovector at the 5\% level and the 5 pc length scale.}}
\label{fig:Bfield_fil}
\end{figure*}

\begin{figure}
\includegraphics[width=\columnwidth]{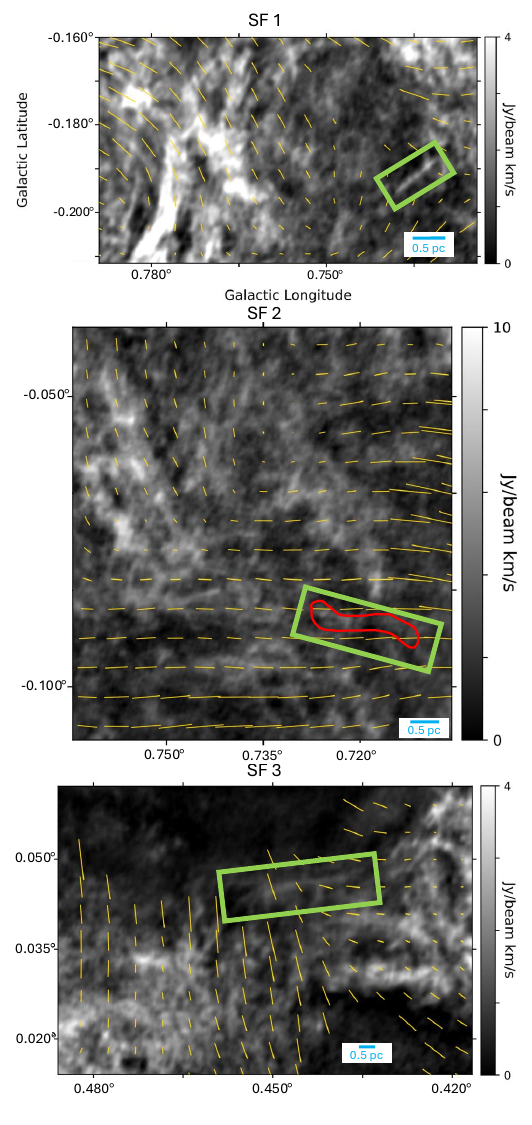}
\caption{Few FIREPLACE POS magnetic field orientations overlap with the three SFs (SF 1 at \textit{top}, SF 2 in the \textit{middle}, and SF 3 on the \textit{bottom}), however those that do show relative angles of about 30\degree~(SF 2) and 90\degree~(SF 3).
\HNCO~moment 0 is shown in grayscale with the magnetic field orientation displayed as yellow line segments. The locations of the SFs within the larger fields of view are indicated with green rectangles. The extent of SF 2  is marked as a red shape since it is not bright in HNCO moment 0, as shown in the middle panel. \textbf{Inset boxes show scale bars of 0.5 pc.}}
\label{fig:Bfield_str}
\end{figure}

We now compare the orientation of the LFs and SFs with the magnetic field orientation derived from FIREPLACE. The magnetic field derived from FIREPLACE is known to largely trace the far-infrared dust emission \citep{Pare2024}. Therefore, in a follow-up work, we compare the FIREPLACE magnetic field orientations with the far-infrared dust emission \citep{Pare2025b}. \updates{We note that the comparison of the alignment (parallel vs. perpendicular) of filament axes and magnetic field directions is subject to significant projection effects. We direct the reader to Sections \ref{sec:morphology_uncertainties}, \ref{sec:Bfield_uncertainties} for more details on these geometric limitations.}

As the first step in our comparison of filament orientation with POS magnetic field orientations, we qualitatively assess whether or not the magnetic field derived from FIREPLACE traces the selected LFs and SFs. Figure \ref{fig:Bfield_fil} presents the \HNCO~moment 0 maps of LF 1 (top), LF 2 (middle), and LF 3 (bottom) with the FIREPLACE pseudovectors overlayed as yellow line segments. We report our overall assessment of the relative magnetic angle for each structure in the last column of Table \ref{tab:filament_striation_statistics}.

In the top panel of Figure \ref{fig:Bfield_fil}, we see that for LF 1, the magnetic field is largely perpendicular to the filament orientation, particularly in regions with lower moment 0 intensity. In regions of higher moment 0 intensity the magnetic field orientation tends to be more parallel and the percentage polarization measurements are generally lower ($<$2\%). However, we observe an enhancement in percentage polarization ($>$5\%) at $\ell=0.5\;$\degree, $b=-0.1\;$\degree\ which is where the moment 0 intensity of the filament is weakest. While there is significant variation in both the percentage polarization and the filament orientation, we conclude that the magnetic field is generally perpendicular to the filament orientation.

LF 2 is shown in the middle panel of Figure \ref{fig:Bfield_fil} which shows that the magnetic field is generally well-aligned (parallel) with this structure. In fact, the magnetic field even appears to curve to follow the filament orientation where the structure exhibits a kink at ($\ell$,$b$) = (0.01\degree, $-$0.01\degree). However, the curvature of the magnetic field is observed throughout the larger field-of-view, so this curvature could be a result of a foreground or large-scale structure (see section \ref{sec:Bfield_uncertainties}).
We see a systematically larger percentage polarization ($p\sim$ 3 -- 5\%) in Northern Galactic latitudes for LF 2 compared to the Southern extent of the structure. In the Southern region the percentage polarization is more generally only $1 - 2$\%.

In the bottom panel of Figure \ref{fig:Bfield_fil}, one can see that the magnetic field local to LF 3 shows a mix of parallel, perpendicular, and other rotations with respect to the filament. The POS magnetic field orientations are parallel to the filament orientation in the Northernmost portion of the filament. Progressing South along the filament length are many regions where the magnetic field is either rotated or perpendicular with respect to the filament orientation. The magnetic field local to LF 3 displays more spatial variation than what is observed for the other two LFs. Percentage polarization is generally constant along the length of LF 3, with values ranging from 1 to 2\%. We do, however, observe larger polarization percentages of $\geq$5\% on either side of the LF. This is especially apparent in the southern-most portion of the LF at $\ell=359.45$\degree. Looking throughout the length of the structure, there does appear to be a channel of decreased percentage polarization that traces the moment 0 intensity of LF 3, similarly to what was found for LF 1.

LF 3 crosses one of the prominent non-thermal filaments (NTFs) that are observed throughout the GC \citep[e.g.,][]{Yusef-Zadeh1987,Yusef-Zadeh2022,Pare2022}. The location and orientation of this NTF is shown as a green rectangle in the bottom panel of Figure \ref{fig:Bfield_fil}.
While we see highly polarized vectors that may be associated with the NTF above and below LF 3, neither the HNCO moment 0 structure nor the FIREPLACE magnetic field pseudovectors that coincide with the LF appear to be  influenced by the presence of this NTF.

Figure \ref{fig:Bfield_str} shows zoom-ins of all three of the chosen SFs. The SFs are small structures, and because of the lower resolution of the FIREPLACE survey (19.6\arcsec) few pseudovectors coincide with the SFs. In fact, there are no pseudovectors coinciding with SF 1 (upper panel of Figure \ref{fig:Bfield_str}). We are therefore unable to evaluate how the magnetic field compares to the orientation of this structure.

SF 2 is shown in the middle panel of Figure \ref{fig:Bfield_str}. This SF is very faint in the larger HNCO moment 0, but is more obvious in the smaller velocity range moment 0 or peak intensity maps (Figures \ref{fig:overview} and \ref{fig:striation_ubiquity}). To more clearly demonstrate its extent on this plot, the outline of SF 2 in peak intensity is shown as a red contour in Figure \ref{fig:Bfield_str}. The orientation of the SF is aligned with the orientation of the rectangular box marking its location in the middle panel of Figure \ref{fig:Bfield_str}. The magnetic field subtends a 30\degree angle with respect to the orientation of this SF.

SF 3 is shown in the bottom panel of Figure \ref{fig:Bfield_str}. We observe roughly four Nyquist-sampled pseudovectors that coincide with this structure. The two Eastern pseudovectors are perpendicular to the SF orientation and the two Western pseudovectors subtend an angle of $\sim$30\degree\ with respect to the SF orientation. There is an apparent gradient of magnetic field orientation where the pseudovectors are increasingly aligned with the structure from East to West. The orientation of these vectors matches the larger-scale field pseudovectors so may be related to larger-scale processes.

\subsubsection{Histogram of relative orientation analysis of magnetic field alignment with filaments}\label{sec:results_bfield_histogram}
We use the PRS method proposed by \cite[][see also Section \ref{sec:method_magnetic}]{Jow2018} to quantitatively analyze the relative orientation between the local intensity structure of the LFs and the magnetic field revealed by the FIREPLACE survey as discussed in Section \ref{sec:method_magnetic}. To match the resolution of SOFIA data, we convolved the integrated intensity (moment 0) maps of HNCO from ACES data to the 19.6\arcsec resolution of the FIREPLACE survey to derive the intensity gradient.

\begin{figure*}
\includegraphics[width=0.48\textwidth]{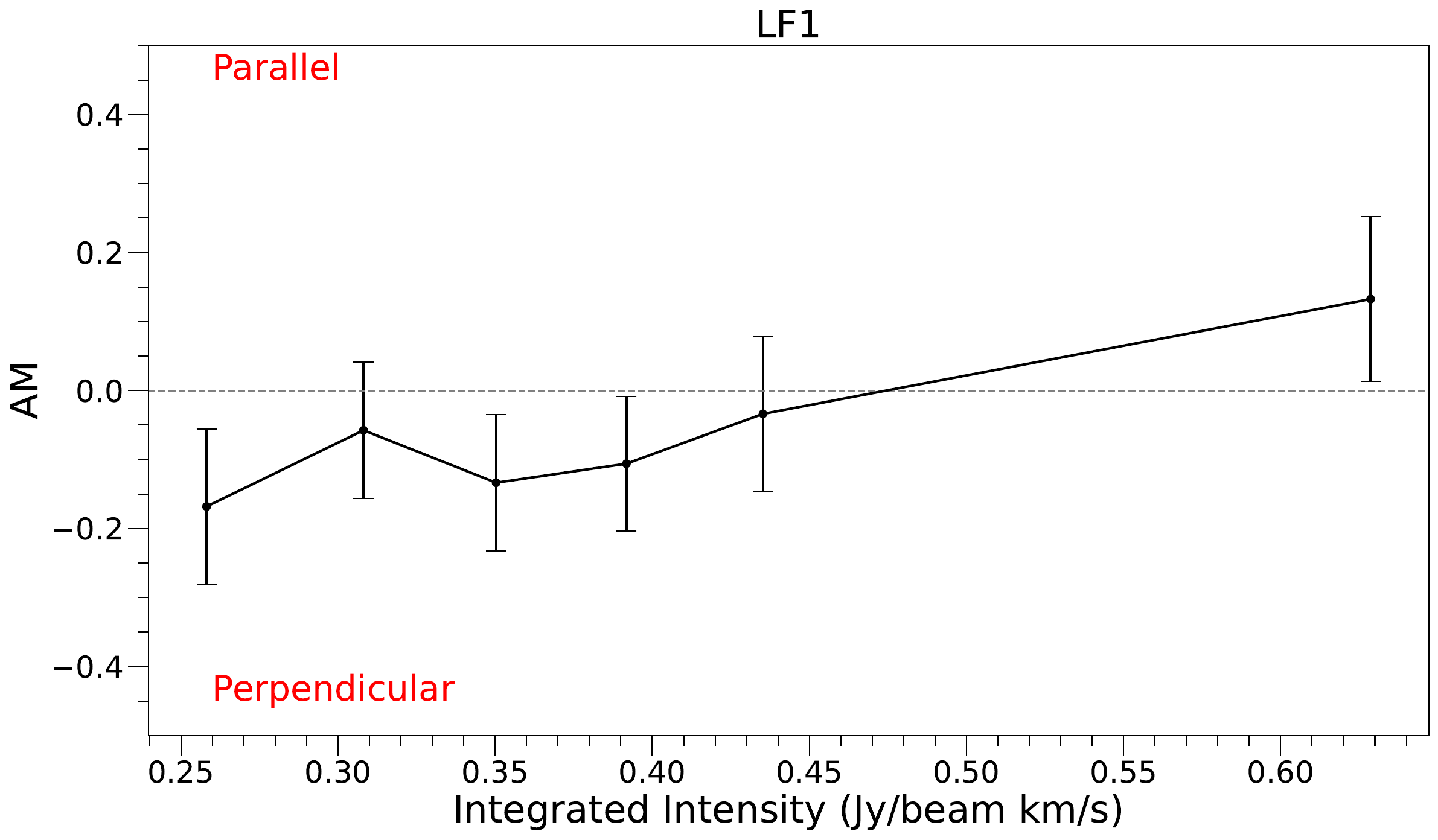}
\includegraphics[width=0.48\textwidth]{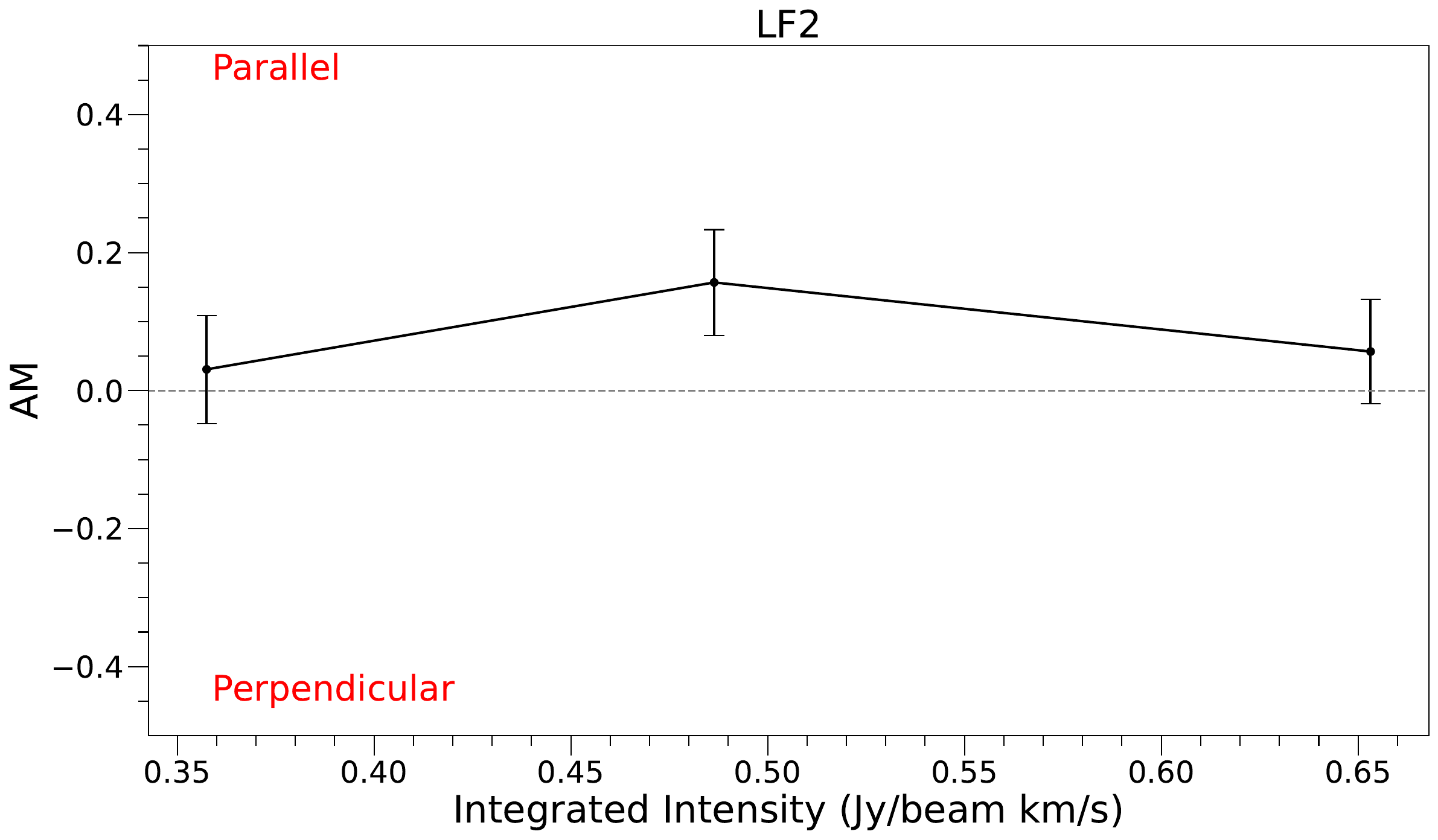}\\
\includegraphics[width=0.48\textwidth]{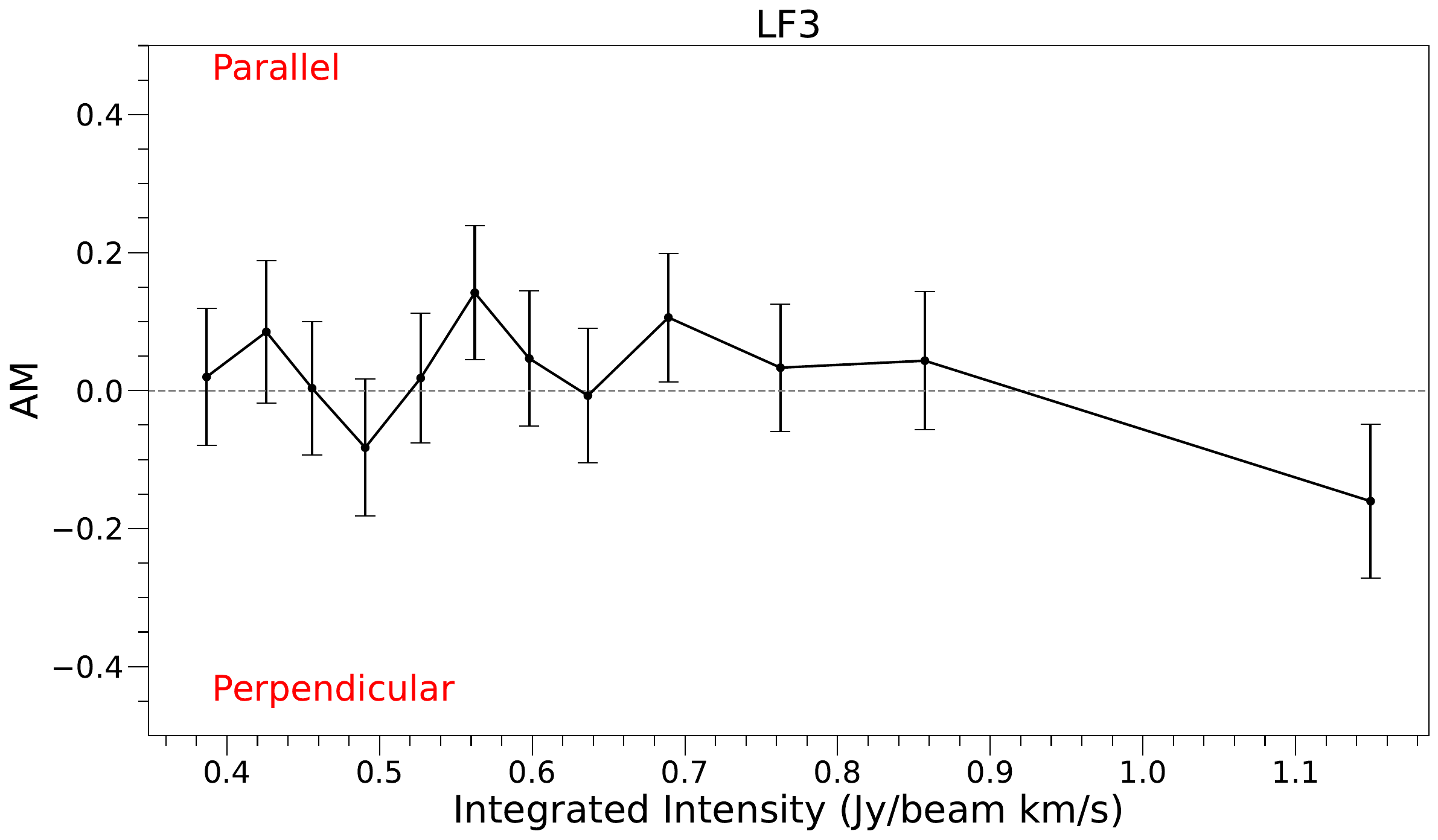}
\includegraphics[width=0.48\textwidth]{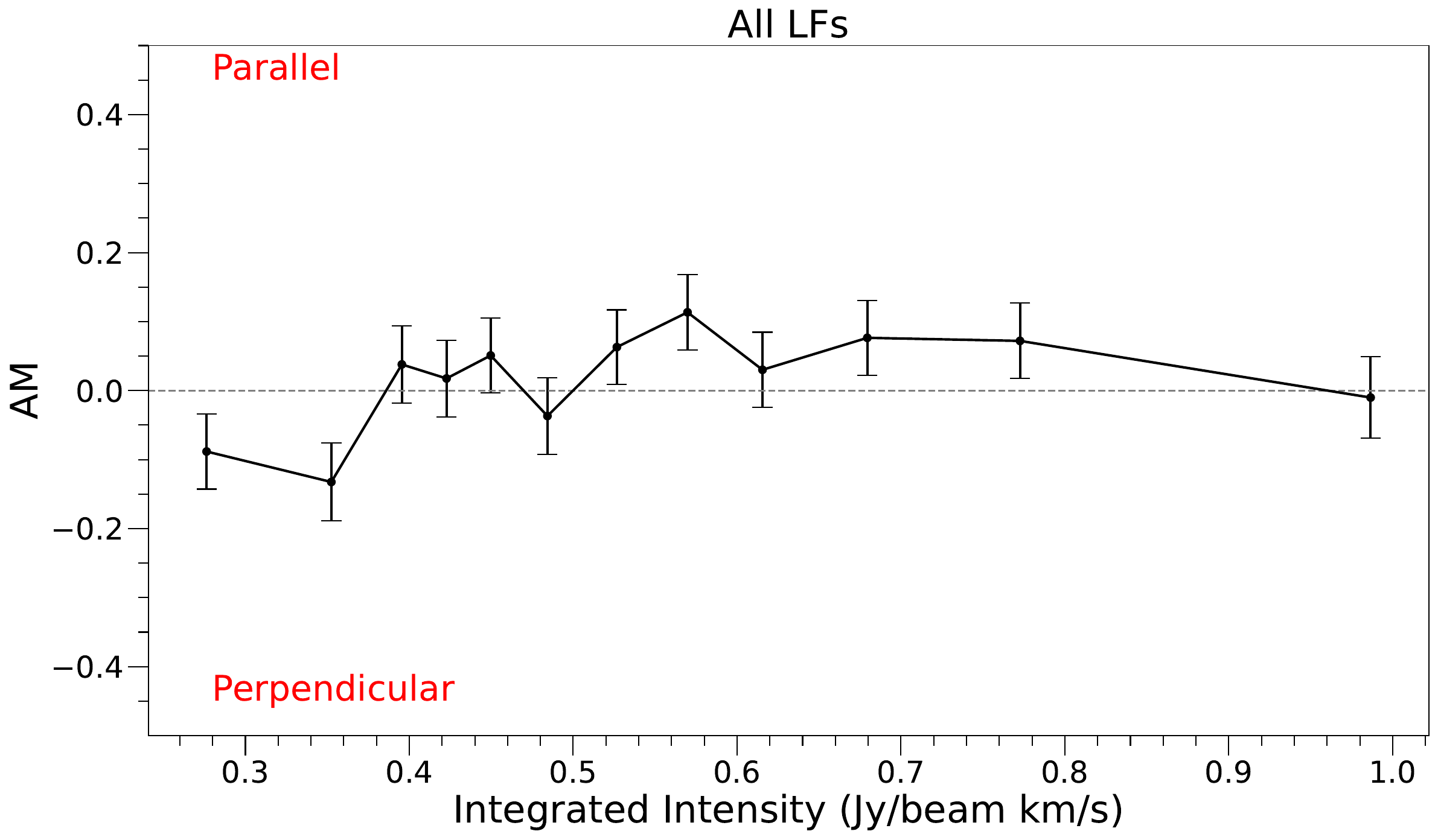}
\caption{A quantitative comparison of structure alignment with magnetic field shows a range of orientations between the LFs and FIREPLACE magnetic field pseudovectors. Plots of the Alignment Measure (AM) calculated for LF 1 (top left), LF 2(top right), and LF 3(bottom left). The bottom right plot displays the average AM values found for the three LFs. The values AM$~>0$ and AM$~<0$ correspond to the magnetic field being oriented preferentially parallel or perpendicular to the iso-intensity contours, respectively.}
\label{fig:HRO}
\end{figure*}

To investigate the trend of Alignment Measure (AM) with intensity, we calculate AM in different intensity bins, each with an equal number of data points. The typical number of pixels per bin are $\sim$100. Since the extent of the SFs are about 2~pc ($\sim50^{\prime\prime}$, at the distance of 8.2~kpc), we don't have adequate POS magnetic field orientations along the SFs for statistical analysis. We therefore only display the AM results for the three larger LFs studied in this work. Figure \ref{fig:HRO} presents the behavior of AM in different intensity bins for each of the three LFs. We also show the average AM over the set of three LFs.

In Figure \ref{fig:HRO} we see a range of AM trends for the different LFs. Here we describe some trends, but first note that most AM values for the three LFs are close to zero. A value of zero would represent a random distribution of relative orientation, and considering the uncertainties, we consider these to be possible, not definitive trends. The quantitative results discussed in this Section are in agreement with the qualitative orientation comparison presented in Section \ref{sec:results_Bfield_vectors}.

LF 1 exhibits negative AM values at lower integrated intensity regimes, indicating a preferentially perpendicular magnetic field orientation compared with intensity structure. At the highest integrated intensity regime, however, the AM transitions to positive indicating a parallel preferential orientation between magnetic field and intensity structure.
The opposite trend is observed for LF 3, where the AM is initially positive at lower integrated intensity regimes and then becomes negative in the highest integrated intensity regime.
Meanwhile, the AM values in LF 2 are positive, meaning that the magnetic field is preferentially aligned with (parallel to) the intensity structures. We determine the average AM values for the range of integrated intensity regimes studied in the individual LFs. The average AM distribution is shown in the lower right panel of Figure \ref{fig:HRO}. The average trend exhibits negative AMs in the lower integrated intensity regimes, and largely positive AMs at higher intensity regimes. The connection between the magnetic field and the LFs is explored in more detail in \citet{Pare2025b} where a larger sample of LFs is studied. For the LFs that overlap with \citet{Pare2025b} we find a $\sim$70\% overlap in filament masks.

\subsection{Comparison with Other Molecular Emission Lines}
\label{sec:chemistry}

We compare the distribution and brightness of the \HNCO~integrated line intensity in the SFs and LFs with other molecular tracers included in ACES:  \SiO, \CS, \SO, \HCtrN, \HthCOp, and \HthCN. Their spectroscopic parameters are summarized in Table~\ref{tab:chemistry_data}.
For the comparison, all data sets were reprojected into a common spatial grid. We then integrated over the velocity extent (Table \ref{tab:filament_striation_statistics}) of each object for each molecular tracer.
In this work, we present comparative figures and discussion for SF 1 only, as it is the most isolated and therefore simplest of the sample. We leave a more detailed analysis of the other structures for follow-up works. We present overall line ratios and correlation coefficients in Table \ref{tab:chemistry_detections} for all three LFs and all three SFs, described later in this section.

\begin{figure*}
\centering
\includegraphics[height=0.86\textheight]{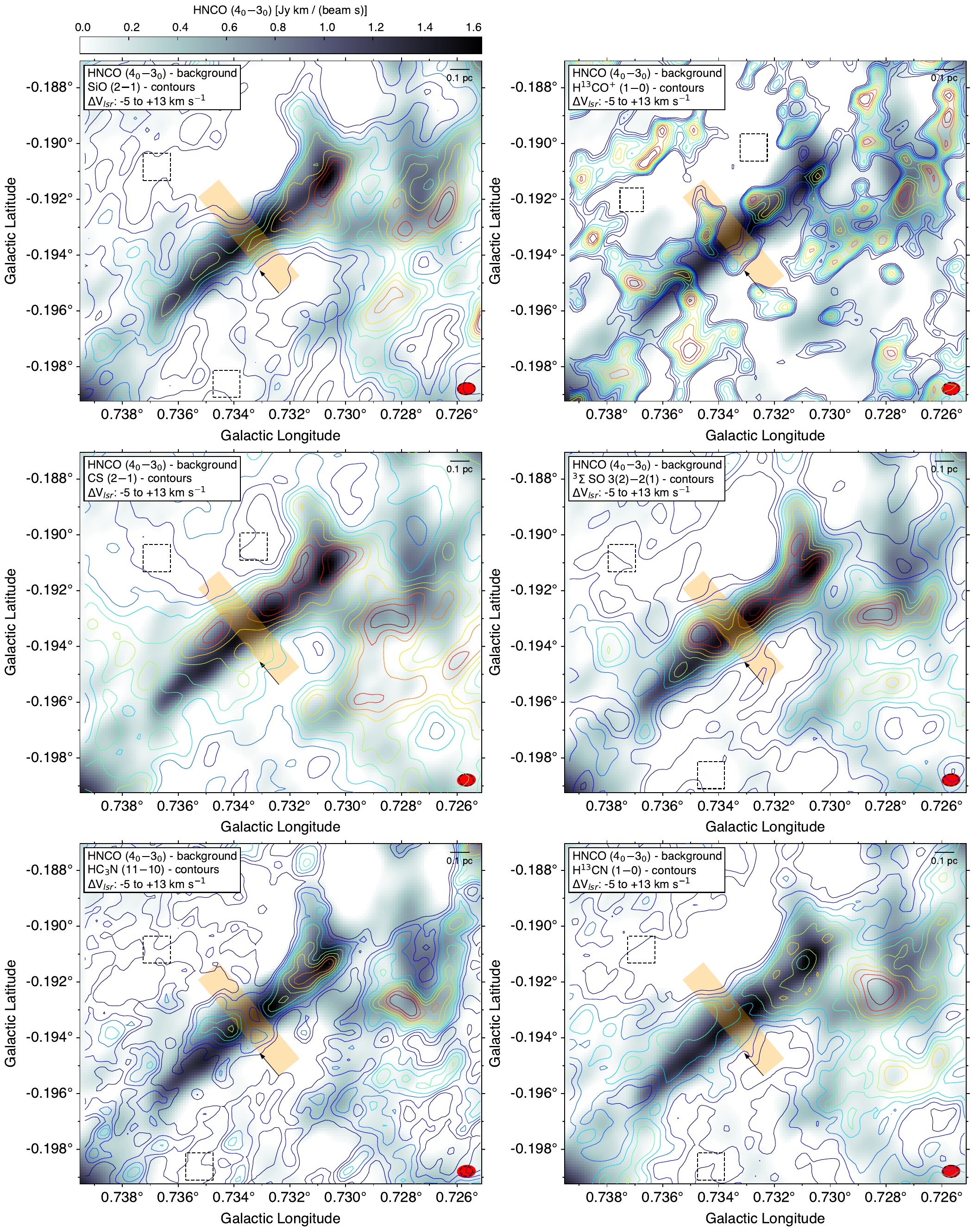}
\caption{A comparison between \HNCO~and a variety of molecular species seen with ACES in SF 1   largely shows good correlation, but with some interesting exceptions and potential trends. Integrated intensity maps of \HNCO~(background) from $-$5 \kms~ to $+$13 \kms~and \SiO, \CS, \HCtrN, \HthCOp, \SO, and \HthCN~(contours) of SF 1. Contour levels are listed as (species: minimum, increment, \# of contours) in the following for their reconstruction in units of Jy beam$^{-1}$   \kms:
(\SiO: 0.016, 0.033, 10);
(\HthCOp: 0.013, 0.007, 10);
(\CS : 0.022, 0.099, 10);
(\SO  : 0.023, 0.047, 10);
(\HCtrN   : 0.012, 0.027, 10); and
(\HthCN : 0.019, 0.034, 10).
The minimum contour in each case is calculated as the average standard deviation between the two low emission regions depicted by the dashed-line black squares. The orange rectangle represents the area used to produce the position-velocity diagrams in Figure \ref{fig:pv_s1_center}, with the origin depicted by the black arrow. The ALMA synthesized beam size is shown as a red ellipse for the \HNCO~transition, and as black open-dashed ellipses for the other species. }
\label{fig:int_map_s1_center}
\end{figure*}

Figure~\ref{fig:int_map_s1_center} compares the spatial distribution of the \HNCO~integrated line intensity (background) with that of the other molecular tracers (colored contours) in ACES for SF 1 . Most  of the molecular emission roughly follows the spatial distribution of the \HNCO~gas, with the exception of the \HthCOp~emission showing a more clumpy distribution across the entire region. The physical conditions along SF 1 seem to vary, with most of the brightest integrated intensities found towards its northern part. The distribution of the \HCtrN~emission tracing warmer gas \citep{Rodriguez-Franco1998,Martin2008, Miettinen2014,Dishoeck1999,Chapman2009,He2021}, with critical density comparable to that of the \HNCO~transition (see Table~\ref{tab:chemistry_detections}), is heavily concentrated towards the north, and almost completely absent in the south, suggesting the presence of a temperature gradient. In its northern part, the brightest \SiO~emission connects, at least in projection, with another structure around $\ell$\;$\sim$\;0.726\degree, $b$\;$\sim$\;$-$0.193\degree~showing similar levels of \SiO~emission, suggesting the presence of shocks. To derive the exact excitation conditions of the gas along SF 1, complementary data (more transitions of the same species) are needed.

\begin{figure*}
\centering
\includegraphics[height=0.9\textheight]{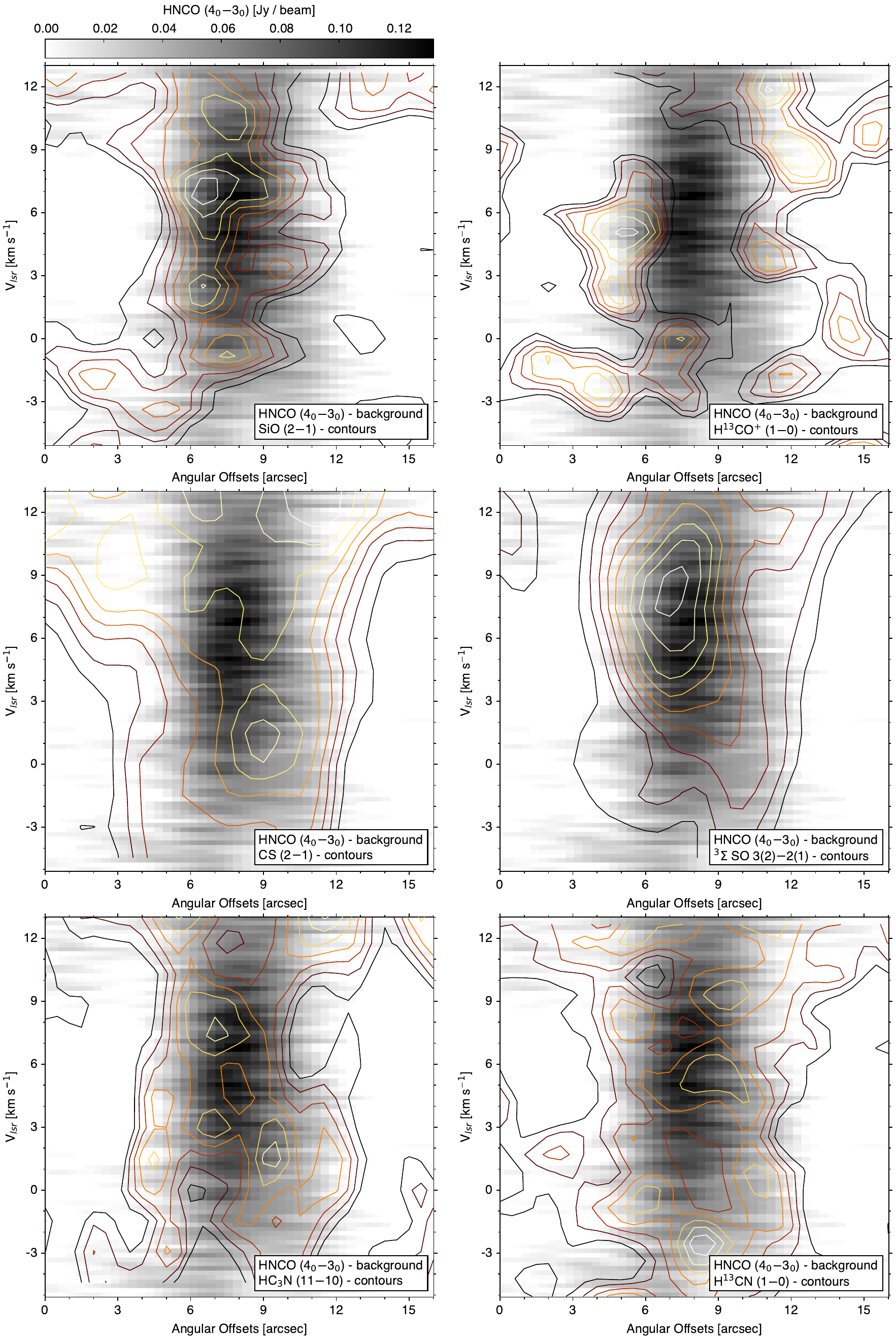}
\caption{A comparison of the \HNCO~position-velocity diagram across SF 1 (from Figure \ref{fig:int_map_s1_center}) with other molecular line species shows that most molecular lines generally follow the \HNCO, however with some interesting deviations. Position-velocity diagrams of \HNCO~(background), \SiO, \CS, \HCtrN, \HthCOp, \SO, and \HthCN~(contours) of SF 1 from the area represented by the orange rectangle in Figure \ref{fig:int_map_s1_center}. Contour levels are listed as (species: minimum, increment, \# of contours) in the following in units of Jy beam$^{-1}$ \kms:
(\SiO: 0.0020, 0.0020, 8);
(\HthCOp: 0.0010, 0.0010, 7);
(\CS: 0.0100, 0.0050, 8);
(\SO: 0.0050, 0.0030, 9);
(\HCtrN: 0.0005, 0.0012, 7); and
(\HthCN: 0.0010, 0.0020, 7).}
\label{fig:pv_s1_center}
\end{figure*}

In order to investigate the kinematics in the central region of SF 1, the orange rectangle (3.5\arcsec$\times$16.5\arcsec) in each panel of Figure~\ref{fig:int_map_s1_center} was used to produce the PV diagrams in Figure \ref{fig:pv_s1_center}. Each PV diagram was generated by integrating the emission of each line within the defined area along the short spatial axis of the rectangle. Each panel of Figure \ref{fig:pv_s1_center} shows \HNCO~in background grayscale and other molecular tracers as colored contours.

In Figure~\ref{fig:pv_s1_center}, the brightest \HNCO~emission (background) is broad along the radial velocity axis, extending from $\sim$+3 \kms~to $\sim$+9 \kms. At the same time, it originates from a narrow angular scale (with a peak between angular offsets of $\sim$6\arcsec~and $\sim$9\arcsec). This is consistent with gas affected by strong turbulence, which are typical conditions of the gas within the CMZ. Most of the molecular gas traced by the rest of the ACES spectral lines (contours) closely follows the \HNCO~emission distribution. In particular, the \SiO, \HCtrN, \HthCOp, and \HthCN~emissions appear quite clumpy in phase-space, while the \CS~emission is the most extended among all observations. Interestingly, the \HthCOp~emission strongly anti-correlates with the \HNCO~emission, surrounding it in phase-space, possibly reflecting different excitation mechanisms or strong opacity effects. The \SO~emission distribution is smooth, resembling a single gas structure, similar to that seen in \SiO, and slightly red-shifted with respect to the \HNCO~bulk emission, suggesting similar excitation mechanisms. No evident large-scale expanding motions \citep[e.g., arc-like signature for bubbles, faint high velocity components for outflows, etc.;][]{Luisi2021,Bonne2022,Lutz2020,Stuber2021} are present in the central part of SF 1, in any of the observed transitions, suggesting that the gas kinematics for SF 1 is likely dominated by the ambient turbulence and not by a local source. A similar analysis for the rest of the objects listed in Table \ref{tab:filament_striation_statistics} is intended for a follow-up publication.

\begin{figure*}
\centering
\includegraphics[width=0.9\textwidth]{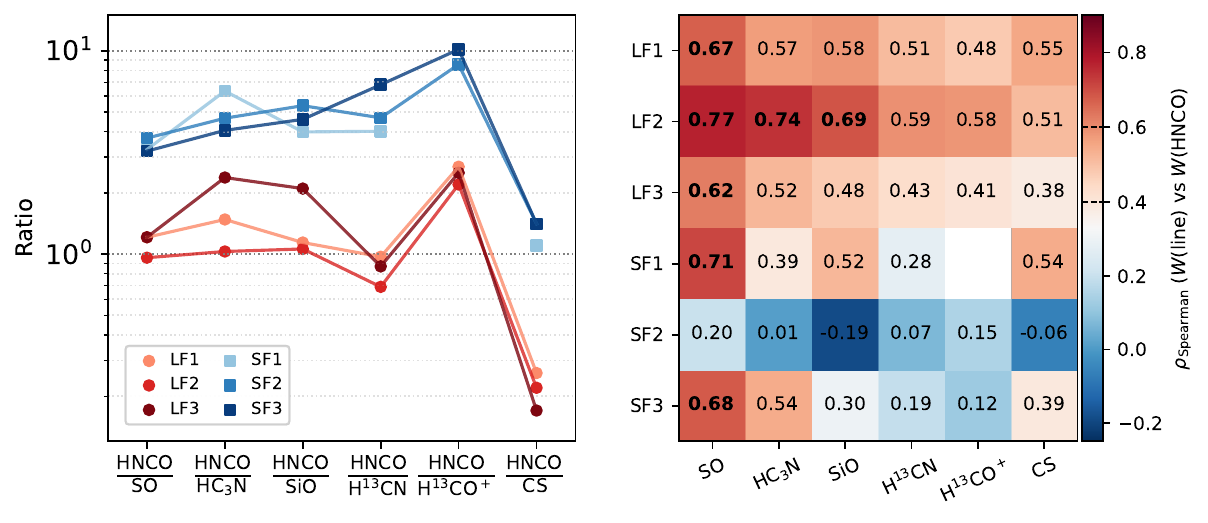}\vspace{-0.4cm}
\caption{For each object,  (left) a comparison of the integrated line intensity (average moment 0, or $W$, in units K~km~s$^{-1}$) ratio (HNCO/line)  and (right)~the~Spearman correlation rank between $W$(line) and $W$(HNCO), for emission within the masked area (see Sections \ref{sec:method_morphology} and \ref{sec:method_kinematic}), with $W$\,$>$\,2\,K\,\kms, \updmay{we note that for SF1, the number of pixels with W(\HthCOp)> 2 K km/s is negligible}.  
Circle \updmay{redish} markers show the ratio values for Large-scale Filamentary structures (LFs), and square \updmay{blueish} markers for Small-scale Filamentary structures (SFs).  The values for the line intensity ratios are tabulated in Table\,\ref{tab:chemistry_detections}.  Correlations with $\rho > 0.6$, in darker red highlight strong correlations, correlations in light orange-red show moderate correlations ($\rho$= [0.4--0.6]), and those with $\rho$ = (-0.2, 0.4), in blue, show that there is no monotonic/linear correlation. }
\label{fig:ratios_corr}
\end{figure*}

\updab{Figure~\ref{fig:ratios_corr}~(left) shows the } integrated line intensity ratios (HNCO/X molecular species) \updab{(see Table\,\ref{tab:chemistry_detections} for the tabulated values)}. \updmay{We order the ratios according to the Spearman correlation rank values for LF2 and LF3 (see Fig.~\ref{fig:ratios_corr} (right)).} We determine the ratios for emission within the masked area (Sections \ref{sec:method_morphology} and \ref{sec:method_kinematic}) \updmay{ where $W$(line) and $W$(\HNCO)>~2~K~\kms}.  \updmay{The lowest ratios for both LFs and SFs are those with \CS, while the highest ratios are those with \HthCOp. We note that for SF1 the number of pixels with W(\HthCOp)> 2 K km/s is negligible so we do not include this in our analysis.} \updab{For every ratio, the SFs show systematically higher values than LFs\updmay{, i.e., the \HNCO\,emission is enhanced in SFs with respect to the other lines, compared to LFs. This suggest that line intensity ratios may help to distinguish the SF and LF populations.}}

\updab{Figure~\ref{fig:ratios_corr} (right)} shows the Spearman's rank correlation coefficients, $\rho$\footnote{We use the Spearman's correlation, instead of the usual Pearson's correlation, because the molecular emission is not normally distributed. This is a non-parametric statistic that measures the monotonic relationship. A value close to 0 does not imply that there is no relationship, just that it is not monotonic.}. 
We find that the strongest correlations are those between \HNCO\;and \SO, for all the structures, followed by \HCtrN~and \SiO.  
\updab{We note that the \HthCOp\, emission in SF1 is specially faint and clumpy, likely noise dominated (see Fig.~\ref{fig:int_map_s1_center}, so we do not include it in the analysis). SF2 is located in the Sgr B2 extended region (see middle panels in Fig.\ref{fig:overview}). The complexity of Sgr B2 may explain the generally low $\rho$ ($\sim$\,0) found for SF2. Interestingly, this object has the largest velocity gradient and the highest median $N$(H$_2$) (see Table\,\ref{tab:filament_striation_statistics}). \updmay{Overall, LFs show higher $\rho$ than SFs, which show a larger scatter and include cases with weak or no correlation. Despite the larger spatial extent of LFs, their molecular emission shows moderate to strong correlations across most lines, suggesting a relatively coherent chemical structure over parsec scales. The correlation ranks between \HNCO\,and \HthCN, and between \HNCO\, and \HthCOp\, are specially useful to differentiate the SF and LF populations. As a summary, this analysis shows that we can differentiate the SF and LF populations using line intensity ratios and the Spearman correlation rank.} }

HNCO is usually a good tracer of low velocity shocks \citep[e.g.,][]{Rod-Fer2010,Kelly2017} and its gas-phase abundance is the result of grain mantle evaporation induced by shock waves. 
Another excellent tracer of (high-velocity, $\sim$\;20$-$50\;\kms) shocks is SiO \citep[e.g.,][]{Martin-Pintado1992}. SiO abundance is specifically enhanced by sputtering of silicate grains, hence, bright SiO emission often traces shocked molecular gas \citep[e.g.,][]{Jimenez-Serra2005,Jimenez2008,Jimenez-Serra2010}. In the Galactic Center, the characteristics (size and H$_2$ densities) of the SiO emission regions are different from those observed in the Galactic disk, interpreted as evidence for large-scale or \updates{global} shocks \citep{Martin-Pintado1997}. These two shock tracers (HNCO and SiO) are chemically different and the line intensity ratio between HNCO and SiO should trace how fast, or strong, the shock is in each region \citep{Kelly2017,He2021, Huang2023}. Therefore, we could explain the higher \HNCO/\SiO\; ratios in the three SFs, compared with the three LFs, if shocks are faster (stronger) in the LFs. If shocks in the SFs 2 and 3 are especially slower (weaker), a lower abundance of SiO is thus expected along the SF, which would also explain the low correlation (Spearman's rank correlation coefficients of -0.19 and 0.30) between \HNCO\;and \SiO\;in these regions. On the other hand, in the LFs, where the \HNCO/\SiO~ratio is lower, the SiO abundance would be higher indicating the presence of higher-velocity shocks \citep{Jimenez2008,Kelly2017} and a better match between these two shock tracers ($\sim$0.5 -- 0.7). \updmay{An in-depth analysis of the chemistry of these objects is intended for a follow-up publication.}

\begin{table*}
\caption{Integrated line intensity ratios with HNCO.}
\label{tab:chemistry_detections}
\begin{tabular}{lcccccc}
\hline
\textbf{Object}  
& \textbf{\SO} 
& \textbf{\HCtrN} 
& \textbf{\SiO} 
& \textbf{\HthCN} 
& \textbf{\HthCOp}  
& \textbf{\CS} \\
\hline
LF 1 & 1.21 & 1.48 & 1.14 & 0.97 & 2.69 & 0.26 \\
LF 2 & 0.96 & 1.03 & 1.06 & 0.69 & 2.20 & 0.22 \\
LF 3 & 1.21 & 2.38 & 2.10 & 0.87 & 2.51 & 0.17 \\
SF 1 & 3.27 & 6.34 & 3.99 & 4.02 & ...  & 1.10 \\
SF 2 & 3.72 & 4.66 & 5.37 & 4.68 & 8.57 & 1.41 \\
SF 3 & 3.21 & 4.06 & 4.60 & 5.81 & 10.16 & 1.41 \\
\hline
\end{tabular}
\medskip
\par\noindent Notes. For each object, integrated line intensity (average moment 0, or $W$, in units K~km~s$^{-1}$) ratio (HNCO/line) for emission within the masked area (see Sections \ref{sec:method_morphology} and \ref{sec:method_kinematic}), with $W$\,$>$\,2\,K\,\kms. \updab{Figure \ref{fig:ratios_corr} (left) summarizes this table.}
\end{table*}

\section{Discussion}
\label{sec:discussion}

\subsection{These Filamentary Structures are Most Likely Located Within the CMZ}
\label{sec:discussion_inthecmz?}

We consider the observed properties of the filamentary structures reported in this work and assess whether or not they are located within the CMZ. The evidence we consider for each structure is: the \HNCO~spectral line width, the presence of shocked gas tracers, and possible association with known CMZ features. Based on this evidence, we assert that it is likely that each of these filamentary structures is located within the CMZ.

The first piece of evidence we evaluate is the \HNCO~spectral linewidth (FWHM, Table \ref{tab:filament_striation_statistics}), which is the average beam-scale linewidth of each structure over its entire mask (see Section \ref{sec:method_kinematic}). Molecular gas in the CMZ tends to be more turbulent on parsec scales than in the disk of the Galaxy by an order of magnitude \citep{Heyer2015, Shetty2012,Kauffmann2017}, meaning that the spectral linewidth is an often-used tool for discerning CMZ vs. foreground objects. The size-linewidth relation in the CMZ follows an approximate trend of $\sigma = \sigma_{0}R^{0.7}$ \citep{Shetty2012} and \citet{Kauffmann2017} reported a best-fitting intercept of $\sigma_{0} = 5.5 \pm 1.0$~\kms. The ACES data has a FWHM spatial resolution of 0.1~pc at the distance of the CMZ\footnote{If one of these objects were foreground, the ACES spatial resolution would correspond to an even smaller physical size, which would imply an even smaller linewidth based on the size-linewidth relationship.}, which we convert to an effective radius \citep[as in][]{Shetty2012} using Equation \ref{eq:reff}, which gives $R_{eff}=0.06 pc$. At this size scale, the characteristic CMZ linewidth based on the above expression is $\sigma = 0.8$~\kms or $\rm FWHM = 1.8$ \kms. In all cases, the average structure FWHM is at least 2$\times$ this value.

Secondly, we consider the detection of molecular line species which trace dense gas, hot gas, and shock chemistry as additional evidence suggesting a likely placement within the CMZ. The filamentary structures were identified as significant detections in \HNCO, which traces low-velocity shocks and dense gas (references in Table \ref{tab:chemistry_data}). The line ratios for each object with six other molecular species are reported in Table \ref{tab:chemistry_detections}. The results in the Table clearly show that these filaments are seen in a wide variety of molecular lines, including \SiO~(tracing high-velocity shocks), \SO~(tracing shocks/dense gas), and \HCtrN~(tracing low velocity shocks/dense gas).

Finally, some of the filamentary structures (LF 1, LF 3, SF 2, and SF 3) have clear associations with known CMZ structures while others have only a tentative association (LF 2 and SF 1). We describe the associations of LF 1, LF 3, SF 2, and SF 3 in the next Section (\ref{sec:connect_3d_structure}), which all appear to be nested within well-known CMZ structures. 
On the other hand, SF 1, while near the M0.8–0.2 ring region \citep{Nonhebel2024}, is much more isolated. However, its average line width ($\rm FWHM = 7.3$ \kms, Table \ref{tab:filament_striation_statistics}) and bright emission across shock-tracing molecular species (Table \ref{tab:chemistry_detections} and Figures \ref{fig:int_map_s1_center} and \ref{fig:pv_s1_center}) suggest that it is still likely within the CMZ. Finally, LF 2 does not lie directly on an orbital stream (Figure \ref{fig:ppv}), however, it may connect the SgrA* region with the orbital stream as an EVF. Additionally, its line width $\rm FWHM = 8.5$ \kms, Table \ref{tab:filament_striation_statistics}) and emission in many molecular species Table \ref{tab:chemistry_detections}) similarly indicate a likely position within the CMZ.

\subsection{Connection with the 3-D structure of the CMZ}
\label{sec:connect_3d_structure}
The motion of the dense gas in the CMZ is often described in terms of either orbital streams or a twisted ellipse on which the bulk of gas and molecular clouds tend to follow on the projected plane of the sky \citep{Walker2025, Lipman2025, Molinari2011, Kruijssen2015,Sofue1995,Sofue2022,Sormani2019, Henshaw2016a}. As reported in Section \ref{sec:kinematics}, LF 1 and LF 3 follow these orbital streams in PPV space. The orbital stream plotted in Figure \ref{fig:ppv} was a by-eye best-fitting ellipse calculated by \citet{Walker2025} based on the results of the 3DCMZ paper series \citep{Battersby2025a, Battersby2025b, Lipman2025, Walker2025}. This paper series focused on dense CMZ gas contained within molecular clouds. However, the correspondence between this orbital model and the extended streams which we identify as LF 1 and LF 3 in this work, in PPV space is encouraging. We therefore suggest that using LFs to help define and fit the 3DCMZ orbital streams may be a  fruitful avenue for pursuit.

LF 1 exists below the Sgr B2 region in PP space, but appears in much higher velocity channels than typically associated with the dust ridge. Clouds along the dust ridge show strong evidence in favor of lying on an orbital stream on the near side of the CMZ in front of SgrA* \citep{Walker2025, Lipman2025, Reid2009, Nogueras2021} The position of LF 1 in PPV space places it near the edge of most proposed geometric orbits, and likely on the far-side of the orbit \citep{Sofue1995, Molinari2011,Kruijssen2015,Walker2025, Lipman2025}. LF 1 also exhibits a similar velocity dispersion and longitudinal length as LF 3, which appears to be associated with the ``wiggles'' region. The wiggles region displays significant quasi-periodic variability in velocity space, possibly driven by gravitational instabilities  towards the edge of the orbital stream \citep{Henshaw2016b, Henshaw2020}. 

As described in some detail in Section \ref{sec:results_PPVspace}, LF 2 is at an intriguing point in PPV space and unraveling the origin of this structure could affect our understanding of accretion towards or outflow from the central region of the Galaxy. LF 2 lies just above the 50 \kms~cloud in PP space, and at an angle of 47\degree, is the most inclined of our filament sample. In velocity, LF 2 is centered at -9.6\kms, about 60 \kms~from the orbital stream (containing LF 1 and the 50 and 20\kms~clouds) closest to it in position. However, its upper velocity extends to 20\kms, nearly connecting the region surrounding SgrA* with this orbital stream. Its large velocity extent (44\kms) over a relatively small longitude extent (about 0.03\degree) would classify it as an EVF \cite[see e.g.][]{Sormani2019b}. Such structures are the topic of an ACES paper in preparation where their nature will be described in more detail (Lipman et al., in prep.).

Both SF 2 and SF 3 appear to be kinematically related to clouds along the `dust ridge.' SF 2 lies deep within a complex filamentary network in the Sgr B2 extended region (see middle panels of Figure \ref{fig:striation_ubiquity}). The nature of this region, and why it is home to this complex filamentary network is not presently known. SF 3 connects dust ridge clouds d and clouds e/f. Morphologically (bottom plot in Figure \ref{fig:striation_ubiquity}), SF 3 appears as a stretched out filament parallel to the Galactic plane and loosely connecting these two clouds, like the stretchy dough remaining as you pull apart two lumps. The broad velocity width in PV space agrees with potential stretching of the clouds along the continuous dust ridge feature.

\subsection{Connection with Inflow, Shear, and `Overshooting' Structures in the CMZ}
The gas structures of the CMZ are strongly influenced by the CMZ's unique place in the Galaxy, and it may be the case that the structures identified in this work emerge from the dynamical interactions between clouds and their environment.

In the previous section (\ref{sec:connect_3d_structure}) we describe that LFs 1 and 3  reside in close correspondence to the bulk gas content of orbital streams seen in Figure~\ref{fig:ppv}. They also lie in proximity to the proposed apocenters of the CMZ streams, sites where collisions may occur with inflowing material \citep{Longmore2013b, Anderson2020,Wallace2022}. Hydrodynamic simulations of these locations in the CMZ orbits have also shown increased shear and turbulence in these locations \citep{Hatchfield2021,Tress2020, Sormani2020}.
In this region, one might expect to see extreme shear in CMZ clouds and violent collisions \citep[e.g.][]{Hatchfield2021, Tress2020} with other gas in the stream, leading to increased density and velocity dispersion. The extreme shear in the structures can also be seen in the molecular line data from Table~\ref{tab:chemistry_data}. LF 2 and 3 both show moderate to strong correlations between the HNCO emission and other shock tracers such as SiO and HC$_3$N. The strong correlations may suggest that these regions represent the locations of the strongest shock interaction in the shear motions experienced by the gas along the orbital streams or in the shocks induced by the accretion or overshooting gas on CMZ clouds.

LF 2 is a particularly interesting source with respect to its PPV structure and chemistry. The structure has strong Spearman correlation coefficients between \HNCO, \SO~(0.65), and \HCtrN~(0.7), indicating the gas is strongly shocked. The kinematics of the filamentary structure also support the idea of LF 2 experiencing strong shear. The PV diagram for this structure in Figure~\ref{fig:statistics_pv} shows a notable peak in brightness at the highest offset distances. This may correspond to a collision between this gas flow and the gas moving independently at its destination. Additionally, as seen in the PPV diagram of Figure~\ref{fig:ppv},
LF 2 shows a sharp increase in dispersion around $\ell = $ 0\degree.The broad and longitude-isolated velocity feature of LF 2 resembles the EVFs identified as stream collision sites with inflowing and overshooting gas material on CMZ orbits by \citet{Sormani2019}, lending credence to the hypothesis that LF 2 represents a similar collision, though perhaps on a different scale.

The broad velocity extent of LF 2, in addition to its position near SgrA* and proximity to the 50 \kms cloud could support the connection of the structure to inflowing material, as the PV structure of the gas seems to show extreme stretch and shear similar to what is seen from inflowing material from the dust lanes onto the CMZ. Because of its velocity structure and line-of-sight position near SgrA* in projection, it appears LF 2 may be a candidate for inflowing material between the CMZ ring and the CND. However, further analysis is required to understand whether this feature is representative of material inflowing from the dust lanes, overshooting material raining back down on the CMZ streams, or gas transport from the 100~pc stream inward towards the CND surrounding Sgr~A*.

\subsection{Insights from Magnetic Fields}
Shocks, turbulence, and magnetic fields are often invoked to explain the existence of large-scale filamentary structures, each of which is dynamically important in the CMZ. According to \cite{2019ApJ...878..157X}, shocks tend to produce filaments perpendicular to a magnetic field, while anisotropic turbulent mixing of gas produces parallel filaments; the ion-neutral decoupling may determine the minimal width of observed filaments. \updates{However, we note that POS filament axes and magnetic field vectors are subject to projection effects, which imposes a significant geometric limitation on the interpretation of magnetic field geometrical association.}

We observe a diversity of configurations in magnetic field alignment across the LFs studied. LF 1 exhibits a predominantly perpendicular magnetic field orientation, LF 2 displays a parallel field, and LF 3 shows a range of alignments.

The perpendicular orientation observed in LF 1 is consistent with expectations for high column density structures within highly supersonic compressible MHD turbulence \citep{2019ApJ...878..157X,2019ApJ...886...17H,2020MNRAS.492..668B}. This trend is also seen through observations of high column structures in the Galactic disk \citep{Soler2019}. However, we note that this perpendicular alignment is not always seen in observational studies of high column density structures in the GC, possibly because of the prevalence of shear in the CMZ \citep{Pare2025}.

Perpendicular or parallel alignment of magnetic field and filamentary structures can  reveal the underlying physical mechanisms. A perpendicular alignment can occur due to shock compression in the presence of a strong magnetic field that primarily compresses gas along the magnetic lines, resulting in structures oriented perpendicular to the magnetic field. Conversely, parallel orientations, like those seen in LF 2 and LF 3, are indicative of MHD turbulence or shear. The underlying dynamics of compressible MHD turbulence involve the perpendicular turbulent mixing of density fluctuations, leading to elongated structures that align with the local magnetic field \citep{2019ApJ...878..157X,2019ApJ...886...17H}. Alternatively, shear can drag the magnetic field in the direction of the cloud orbital motion, leading to a parallel alignment. 

\subsection{Known Uncertainties}

\subsubsection{Limitations of morphological analysis}\label{sec:morphology_uncertainties}
This work focuses on presentation of a small representative sample of LFs and SFs. There are uncertainties in both the identification of the structures as well as in the derivation of their properties. We refer the reader to the ACES paper series \citep{Longmore2025, Ginsburg2025, Walker2025b, Lu2025, Hsieh2025} for details on the statistical noise in the ACES data.

In a follow-up paper \citep{Pare2025b} we use the RHT method to identify a larger sample of LFs. In that work, we identify 12 LFs total and recover the three reported here with high fidelity. Unsurprisingly, the exact boundaries of our structures and those using the method from \citet{Pare2025b} differ slightly, but they largely trace the same structure (70\% overlap). Structure identification in the hierarchical ISM is a perpetual challenge as structures do not have clean boundaries, nor are they inherently well-defined. Properties derived will depend upon the classification methods imposed. In this work, we focused on the quantitative analysis of structures that were initially identified by our team, but in the future, hope to implement survey-wide filament identification to the ACES data.

All of the physical properties (Table \ref{tab:filament_striation_statistics}) derived in this work depend not only upon the defined structure boundaries, which are themselves uncertain as noted above, but also on the systematic errors. While we describe some statistical errors in Section \ref{sec:method_morphology}, there are systematic errors of unknown magnitude. Dust opacity, temperature, flux, and distance uncertainties result in a factor of two uncertainty in the calculation of masses in Monte Carlo simulations from \citet{Battersby2010}. Additionally, \HNCO is just one molecular line tracer and has its own abundance and excitation uncertainties. Due to similar uncertainties as tested in \citet{Battersby2010} we expect that the systematic errors in our analysis are of a similar magnitude and outweigh the reported statistical errors.

For example, the widths of the filaments are uncertain and tracer-dependent. The ACES 3\arcsec~beam is about 0.1 pc on the sky at a Galactic Center distance of 8.2 kpc, which means that our filament widths are between slightly less than 1 beam and 7 beams wide. Since we perform a full profile fit \citep[using RadFil][]{Zucker2018_2} we can centroid with higher accuracy than a beam width. These widths are calculated at every pixel along the spine of the filament, so the single ``width" reported actually corresponds to an average across the entire length. The filament widths we report in Table \ref{tab:filament_striation_statistics} are deconvolved from the beam as described in Section \ref{sec:method_morphology} with an uncertainty of 0.07~pc as determined by the standard deviation of the widths. However, the properties are dependent upon our dataset and choice of \HNCO, and would vary depending on the tracer and resolution of the dataset. Because the structure widths are close to the beam size, they may be narrower in higher resolution data and would likely vary depending on molecular line tracer observed (see e.g. Figure \ref{fig:int_map_s1_center}).

It should be noted, as referenced in Section \ref{sec:method_morphology}, that by measuring the morphology of SFs and LFs via the available POS projections, we lack information regarding the radial dimension of observation. Specifically, any structure aligned in the radial dimension along the line of sight will be subject to foreshortening, resulting in an underestimation of the length. Additionally, a structure aligned with the radial dimension may appear to have a stronger curvature, whereas a structure aligned along the line of sight projection may appear more linear.

Assuming a mean orientation of LFs and SFs with respect to the CMZ along the projected line of sight, one would expect the lack of radial position information to give rise to a systematic overestimation or underestimation of curvature in projected ($\ell$,$b$) space. In other words, if LFs and SFs were found to be largely perpendicular to their orbits around the CMZ, we would find shorter measurements and higher curvatures around central longitudes and longer measurements and linear curvatures around outer longitudes. Conversely, if LFs and SFs were found to occur in more parallel orientation with respect to their orbits, we would expect to see shorter, more curved structures at outer longitudes and longer, more linear structures at central longitudes. Though the data we have analyzed in this paper is insufficient in sample size to draw specific conclusions, we highlight this area of uncertainty here insofar as it may prove relevant to further work. In particular, the orientation of these structures could be used to better trace the morphology of 3-D kinematic orbital streams (as presented in Section \ref{sec:connect_3d_structure}).

\subsubsection{Potential line of sight magnetic field contributions}\label{sec:Bfield_uncertainties}

It is important to consider that the magnetic field inferred from the FIREPLACE survey may originate from foreground, magnetized structures. The FIREPLACE team concluded that the field coinciding with the prominent CMZ clouds in high $I_{214}$ emission was likely local to the CMZ \citep{Pare2024}. The LFs and SFs observed in this work, however, do not coincide with high $I_{214}$ emission. We therefore cannot rule out the possibility that the magnetic field is tracing a different magnetized structure located along the line of sight. However, we note that since \HNCO~traces high density gas and FIREPLACE POS magnetic field orientations are density-weighted along the line of sight, it is likely that we are tracing the same structures.

To help evaluate this possibility, we return to Figure \ref{fig:Bfield_fil}. Trends in the magnetic field shown in the top and middle panels of Figure \ref{fig:Bfield_fil} are also observed in surrounding pseudovectors that do not coincide with the filamentary structures. For example, the observed curvature in the magnetic field coinciding with LF 2 in Figure \ref{fig:Bfield_fil}, which seems to follow the bend of the filament, is observed throughout the field of view of Figure \ref{fig:Bfield_fil}. This curve in the magnetic field could therefore be tracing a larger structure.
However, the percentage polarization and magnetic field orientation in the LF 3 region do seem to be influenced by the presence of this structure, as can be seen in the bottom panel of Figure \ref{fig:Bfield_fil}. Percentage polarization coinciding with the filament is a few percent lower than surrounding pseudovectors. This behavior could indicate that the FIREPLACE magnetic field local to this LF is tracing the filamentary structure rather than some other magnetized structure.

Follow-up polarimetric observations with higher spatial resolutions  help resolve this potential complication. The ACES ALMA observations have an order of magnitude improved angular resolution compared to the FIREPLACE observations. Polarimetric ALMA observations with a similar resolution will yield additional pseudovectors coinciding with the LFs and SFs. The larger number of pseudovectors will improve our ability to assess the relative alignment of the magnetic both by-eye and using quantitative methods like the PRS. In \citet{Pare2025b}, we perform a more comprehensive comparison with magnetic fields using a larger sample of LFs.

\section{Conclusions}
\label{sec:conclusions}
New data from the ALMA large program ACES (ALMA CMZ Exploration Survey) reveal \updates{omnipresent} filamentary structure in the molecular gas of the Central Molecular Zone (CMZ) on scales from tenths to tens of parsecs. In this paper, we focus on the \HNCO~molecular line, generally thought to trace dense gas and low-velocity shocks. We demonstrate the preponderance of filamentary structures in the CMZ and quantitatively analyze a small \updates{representative} sample of them. Within the CMZ, we identify two main categories of filaments: Large-scale Filamentary Structures (LFs, with lengths on order of $\sim$10~pc) and Small-scale Filamentary Structures (SFs, with lengths on order of $\sim$1~pc). There may be a yet unexplored distribution of structures between the two identified here, and \updates{we do not identify a complete sample of structures nor demarcate definitive boundaries of these classifications.} We \updates{identify} three example structures in each category and present quantitative measurements towards these six objects. Below we summarize the main conclusions of this work and outline potential directions for future research.

\begin{itemize}
  \item The three LFs we study have 2D projected lengths of 32~pc, 9.8~pc, 49~pc, widths (FWHM) of 0.16~pc, 0.18~pc, 0.69~pc, and position angles with respect to the Galactic plane of 7\textdegree, 47\textdegree, and 27\textdegree. They are largely coherent in velocity space with a few gaps. The line widths (FWHM) of the LFs are 4.7, 8.5, 4.3~\kms.
  \item The three SFs we study have 2D projected lengths of 1.3~pc, 2.1~pc, 3.4~pc, widths (FWHM) of 0.08~pc, 0.12~pc, 0.14~pc, and position angles with respect to the Galactic plane of 43\textdegree, 29\textdegree, and 18\textdegree. They are coherent in velocity space and have line widths (FWHM) of 7.3, 4.4, and 6.9~\kms.
  \item Two of the LFs (LF 1 and LF3) clearly trace the CMZ orbital streams in PPV space, making such structures potentially useful for studies of the 3D structure and mass flows of gas in the CMZ.
  \item LF 2 matches the properties (compact structure, high velocity extent) of an Extended Velocity Feature (EVF). It appears as though LF 2 may connect the SgrA* region with the orbital streams, making it an excellent candidate for gas inflow to the central region and worthy of further study.
  \item Two of the SFs (SF 2 and SF 3) are part of large, complex filamentary networks of similarly-sized filamentary structures. SF 1 was chosen for its relative isolation, but otherwise appears similar to the other SFs.
  \item We investigate the alignment between the LFs and SFs with Plane-of-Sky (POS) magnetic field orientations from the FIREPLACE (Far-InfraREd Polarimetric Large Area CMZ Exploration) survey. We perform a quantitative analysis of the magnetic alignment with the LFs and find that one is largely perpendicular, another parallel, and the third mixed. However, the alignment varies across the LFs and as a function of HNCO intensity, and there seems to be no clear relation between the magnetic field and the high density portions of the filaments. The SFs are analyzed qualitatively, but the overlapping vectors are minimal.
  \item We compute line intensity ratios and Spearman correlation coefficients between the \HNCO~maps and a variety of other molecular lines (see Fig.\,\ref{fig:ratios_corr} and Table \ref{tab:chemistry_data}) and overplot the moment 0 maps for SF 1. \updab{ \SO\;(\HCtrN\ and \SiO\ )} usually show strong (moderate) correlation with \HNCO, while \HthCOp\ and \HthCN\ show \updab{weaker} correlation. There is variety in the correlations across both the LF and SF samples. \updab{The line intensity ratios HNCO/line \updmay{and correlation ranks} successfully differentiate the LF and SF populations.}  Initial investigation in the morphology of the different tracers suggests that further analysis of the chemical signatures of these structures will provide critical information in decoding their physical origin.
\end{itemize}

In this work, we present a narrative of \updates{predominantly}  filamentary structure in CMZ molecular gas as well as a quantitative study of a sample of three objects in each of two main classes of CMZ filamentary structures (SFs and LFs). Follow-up works study a larger sample of EVFs in the CMZ (Lipman et al., in prep.) and the connection between a larger sample of LFs and POS magnetic field orientations \citep{Pare2025b}. Identifying a larger sample of both LFs  and SFs  will be critical in determining their origins. And finally, comparing the emission from different molecular lines has the promise to provide clues into their formation mechanisms.

\section*{Software}
The work completed in this paper made extensive use of the following open-source astronomy software packages: astropy \citep{astropy:2013, astropy:2018, astropy:2022}, spectral-cube \citep{Ginsburg2019spectralcube}, pvextractor \citep{Ginsburg2016pvextractor}, reproject \citep{Robitaille2020reproject},  FilFinder \citep{Koch2015},  RadFil \citep{Zucker2018_2}, glue \citep{Goodman2012, Beaumont2015, Robitaille2019}, numpy \citep{2020NumPy-Array}, and matplotlib \citep{hunter2007matplotlib}. This project also made use of \texttt{CARTA} for data visualisation and exploration \citep{carta}.

\section{Data Availability}
All data products and associated documentation can be found at \url{https://almascience.org/alma-data/lp/aces}. To maximise usability and provide a single, comprehensive resource for users, we also provide an online, machine-readable table, which contains the full, detailed list of all data products released for the entire survey, including hyperlinks to the full files from the ALMA archive.

All code and data processing issues are available at the public GitHub repository here: \url{https://github.com/ACES-CMZ/reduction_ACES}.

The ACES data reduction was a monolithic work that was written up in 5 papers.  If you use the ACES data, please cite the appropriate works, which includes \citet{Longmore2025} and the data papers: continuum \citep{Ginsburg2025} and cubes in high-resolution \citep{Walker2025b}, medium-resolution \citep{Lu2025}, and low-resolution \citep{Hsieh2025} spectral windows. Note that all papers should be cited for use of any data; the continuum imaging relied on the line papers, and vice-versa.

\section*{Acknowledgements}

\updates{The authors thank the referee and scientific editor at MNRAS, whose thoughtful feedback has improved this manuscript. }

Author Contribution Statement: This work was written as part of a ``paper sprint," in which a dedicated team of researchers engaged in an intense, two-week collaborative research and writing process. Please reach out to Cara Battersby if you would like more details on the organization of the paper sprint.
The members of the paper sprint team are the first 17 authors of the paper, followed by members of the data team, project management team, and then additional contributing members of the ACES collaboration. In addition, we thank Brendan DuBois for his guidance on organizing the paper sprint and for his contributions to the project.

This paper makes use of the following ALMA data: ADS/JAO.ALMA\#2021.1.00172.L. ALMA is a partnership of ESO (representing its member states), NSF (USA) and NINS (Japan), together with NRC (Canada), NSTC and ASIAA (Taiwan), and KASI (Republic of Korea), in cooperation with the Republic of Chile. The Joint ALMA Observatory is operated by ESO, AUI/NRAO and NAOJ. The National Radio Astronomy Observatory is a facility of the National Science Foundation operated under cooperative agreement by Associated Universities, Inc.

C.B.\   gratefully  acknowledges  funding  from  National  Science  Foundation  under  Award  Nos. 2108938, 2206510, and CAREER 2145689, as well as from the National Aeronautics and Space Administration through the Astrophysics Data Analysis Program under Award ``3-D MC: Mapping Circumnuclear Molecular Clouds from X-ray to Radio,” Grant No. 80NSSC22K1125.

J.W. gratefully acknowledges funding from National Science Foundation under Award Nos. 2108938 and 2206510.

M.G.S.M.\ acknowledges support from the NSF under grant CAREER 2142300. M.G.S.M.\ also thanks the Spanish MICINN for funding support under grant PID2023-146667NB-I00.

X.L.\ acknowledges support from the Strategic Priority Research Program of the Chinese Academy of Sciences (CAS) Grant No.\ XDB0800300, the National Key R\&D Program of China (No.\ 2022YFA1603101), State Key Laboratory of Radio Astronomy and Technology (CAS), the National Natural Science Foundation of China (NSFC) through grant Nos.\ 12273090 and 12322305, the Natural Science Foundation of Shanghai (No.\ 23ZR1482100), and the CAS ``Light of West China'' Program No.\ xbzg-zdsys-202212.

X.P.\ is supported by the Smithsonian Astrophysical Observatory (SAO) Predoctoral Fellowship Program. Q.Z. acknowledges the support from the National Science Foundation under Award No. 2206512.

Y.H.\ acknowledges the support for this work provided by NASA through the NASA Hubble Fellowship grant No.~HST-HF2-51557.001 awarded by the Space Telescope Science Institute, which is operated by the Association of Universities for Research in Astronomy, Incorporated, under NASA contract NAS5-26555.

A.L.\ acknowledges the support of NSF grants AST 2307840 and ALMA SOSPADA-016.

C.F.\ acknowledges funding provided by the Australian Research Council (Discovery Project grants~DP230102280 and~DP250101526), and the Australia-Germany Joint Research Cooperation Scheme (UA-DAAD).

I.J-.S, L.C. and V.M.R. acknowledge support from the grant PID2022-136814NB-I00 by the Spanish Ministry of Science, Innovation and Universities/State Agency of Research MICIU/AEI/10.13039/501100011033 and by ERDF, UE. \updates{I.J.-S. also acknowledges support from the ERC grant OPENS, GA No. 101125858, funded by the European Union. The project that gave rise to these results received the support of a fellowship from the ”la Caixa” Foundation (ID 100010434)”. The
fellowship code is LCF/BQ/PR25/12110012.}

V.M.R. also acknowledges the grant RYC2020-029387-I funded by MICIU/AEI/10.13039/501100011033 and by "ESF, Investing in your future", and from the Consejo Superior de Investigaciones Cient{\'i}ficas (CSIC) and the Centro de Astrobiolog{\'i}a (CAB) through the project 20225AT015 (Proyectos intramurales especiales del CSIC); and from the grant CNS2023-144464 funded by MICIU/AEI/10.13039/501100011033 and by “European Union NextGenerationEU/PRTR”.

R.S.K.\ and S.C.O.G.\ acknowledge financial support from the European Research Council via the ERC Synergy Grant ``ECOGAL'' (project ID 855130),  from the German Excellence Strategy via the Heidelberg Cluster of Excellence (EXC 2181 - 390900948) ``STRUCTURES'', and from the German Ministry for Economic Affairs and Climate Action in project ``MAINN'' (funding ID 50OO2206).

R.S.K.\ also thanks the Harvard-Smithsonian Center for Astrophysics and the Radcliffe Institute for Advanced Studies for their hospitality during his sabbatical, and the 2024/25 Class of Radcliffe Fellows for highly interesting and stimulating discussions.

M.C.S.\ acknowledges financial support from the European Research Council under the ERC Starting Grant ``GalFlow'' (grant 101116226) and from Fondazione Cariplo under the grant ERC attrattivit\`{a} n. 2023-3014

A.S.-M.\ acknowledges support from the RyC2021-032892-I grant funded by MCIN/AEI/10.13039/501100011033 and by the European Union `Next GenerationEU’/PRTR, as well as the program Unidad de Excelencia Mar\'ia de Maeztu CEX2020-001058-M, and support from the PID2020-117710GB-I00 (MCI-AEI-FEDER, UE).

J.K.\ is supported by the Royal Society under grant number RF\textbackslash ERE\textbackslash231132, as part of project URF\textbackslash R1\textbackslash211322.

J.E.P.\ acknowledges support from the Max-Planck Society.

P.G.\ is sponsored by the Chinese Academy of Sciences (CAS), through a grant to the CAS South America Center for Astronomy (CASSACA). He acknowledges support by the China-Chile Joint Research Fund (CCJRF No. 2312). CCJRF is provided by Chinese Academy of Sciences South America Center for Astronomy (CASSACA) and established by National Astronomical Observatories, Chinese Academy of Sciences (NAOC) and Chilean Astronomy Society (SOCHIAS) to support China-Chile collaborations in astronomy.

\updates{F.N.L.\ gratefully acknowledges financial support from grant PID2024-162148NA-I00, funded by MCIN/AEI/10.13039/501100011033 and the European Regional Development Fund (ERDF) “A way of making Europe”, from the Ramón y Cajal programme (RYC2023-044924-I) funded by MCIN/AEI/10.13039/501100011033 and FSE+, and from the Severo Ochoa grant CEX2021-001131-S, funded by MCIN/AEI/10.13039/501100011033.}

The National Radio Astronomy Observatory and Green Bank Observatory are facilities of the U.S. National
Science Foundation operated under cooperative agreement by Associated Universities, Inc.

\bibliographystyle{mnras}
\bibliography{refs_3dcmz}

\section*{Affiliations}
\printaffiliation{uconn}{Department of Physics, University of Connecticut, 196A Auditorium Road, Unit 3046, Storrs, CT 06269, USA}
\printaffiliation{uflorida}{Department of Astronomy, University of Florida, P.O. Box 112055, Gainesville, FL 32611, USA}
\printaffiliation{iff}{Instituto de Física Fundamental (CSIC). Calle Serrano 121-123, 28006, Madrid, Spain}
\printaffiliation{villanova}{Department of Physics, Villanova University, 800 E. Lancaster Ave., Villanova, PA 19085, USA}
\printaffiliation{cassaca}{Chinese Academy of Sciences South America Center for Astronomy, National Astronomical Observatories, CAS, Beijing 100101, China}
\printaffiliation{ucn}{Instituto de Astronom\'ia, Universidad Cat\'olica del Norte, Av. Angamos 0610, Antofagasta, Chile}
\printaffiliation{cab_csic}{Centro de Astrobiología (CAB), CSIC-INTA, Carretera de Ajalvir km 4, Torrejón de Ardoz, 28850 Madrid, Spain}
\printaffiliation{nanjing}{School of Astronomy and Space Science, Nanjing University, 163 Xianlin Avenue, Nanjing 210023, P.R.China}
\printaffiliation{nanjing_key}{Key Laboratory of Modern Astronomy and Astrophysics (Nanjing University), Ministry of Education, Nanjing 210023, P.R.China}
\printaffiliation{cfa}{Center for Astrophysics | Harvard \& Smithsonian, 60 Garden Street, Cambridge, MA, 02138, USA}
\printaffiliation{ukarcnode}{UK ALMA Regional Centre Node, Jodrell Bank Centre for Astrophysics, Oxford Road, The University of Manchester, Manchester M13 9PL, United Kingdom}
\printaffiliation{jpl}{Jet Propulsion Laboratory, California Institute of Technology, 4800 Oak Grove Drive, Pasadena, CA, 91109, USA}
\printaffiliation{ias_hubble}{Institute for Advanced Study, 1 Einstein Drive, Princeton, NJ 08540, USA (NASA Hubble Fellow)}
\printaffiliation{uw_madison}{Department of Astronomy, University of Wisconsin-Madison, Madison, WI 53706, USA}
\printaffiliation{shao}{Shanghai Astronomical Observatory, Chinese Academy of Sciences, 80 Nandan Road, Shanghai 200030, P. R. China}
\printaffiliation{naoc_key}{State Key Laboratory of Radio Astronomy and Technology, A20 Datun Road, Chaoyang District, Beijing, 100101, P. R. China}
\printaffiliation{kansas}{Department of Physics and Astronomy, University of Kansas, 1251 Wescoe Hall Drive, Lawrence, KS 66045, USA}
\printaffiliation{eso}{European Southern Observatory (ESO), Karl-Schwarzschild-Stra{\ss}e 2, 85748 Garching, Germany}
\printaffiliation{naoj}{National Astronomical Observatory of Japan, 2-21-1 Osawa, Mitaka, Tokyo 181-8588, Japan}
\printaffiliation{ljmu}{Astrophysics Research Institute, Liverpool John Moores University, IC2, Liverpool Science Park, 146 Brownlow Hill, Liverpool L3 5RF, UK}
\printaffiliation{mpia}{Max Planck Institute for Astronomy, K\"{o}nigstuhl 17, D-69117 Heidelberg, Germany}
\printaffiliation{cool}{COOL Research DAO}
\printaffiliation{colorado}{Center for Astrophysics and Space Astronomy, Department of Astrophysical and Planetary Sciences, University of Colorado, Boulder, CO 80389, USA}
\printaffiliation{iaa_taipei}{Institute of Astronomy and Astrophysics, Academia Sinica, 11F of ASMAB, AS/NTU No. 1, Sec. 4, Roosevelt Road, Taipei 10617, Taiwan}
\printaffiliation{eao}{East Asian Observatory, 660 N. A'ohoku, Hilo, Hawaii, HI 96720, USA}
\printaffiliation{chalmers}{Department of Space, Earth and Environment, Chalmers University of Technology, SE-412 96 Gothenburg, Sweden}
\printaffiliation{surrey}{Department of Physics, University of Surrey, Guildford GU2 7XH, UK}
\printaffiliation{insubria}{Universit{\`a} dell’Insubria, via Valleggio 11, 22100 Como, Italy}
\printaffiliation{nrao}{National Radio Astronomy Observatory, 520 Edgemont Road, Charlottesville, VA 22903, USA}
\printaffiliation{anu}{Research School of Astronomy and Astrophysics, Australian National University, Canberra, ACT 2611, Australia}
\printaffiliation{ita_heidelberg}{Universit\"{a}t Heidelberg, Zentrum f\"{u}r Astronomie, Institut f\"{u}r Theoretische Astrophysik, Albert-Ueberle-Str. 2, 69120 Heidelberg, Germany}
\printaffiliation{northwestern_ciera}{Center for Interdisciplinary Exploration and Research in Astrophysics (CIERA), Northwestern University, Evanston, IL 60208, USA}
\printaffiliation{mpir}{Max-Planck-Institut f{\"u}r Radioastronomie, Auf dem H{\"u}gel 69, 53121 Bonn, Germany}
\printaffiliation{ucl}{Department of Physics and Astronomy, University College London, Gower Street, London WC1E 6BT, United Kingdom}
\printaffiliation{izw_heidelberg}{Universit\"{a}t Heidelberg, Interdisziplin\"{a}res Zentrum f\"{u}r Wissenschaftliches Rechnen, Im Neuenheimer Feld 225, 69120 Heidelberg, Germany}
\printaffiliation{radcliffe}{Radcliffe Institute for Advanced Studies at Harvard University, 10 Garden Street, Cambridge, MA 02138, U.S.A.}
\printaffiliation{eso_chile}{European Southern Observatory, Alonso de C\'ordova, 3107, Vitacura, Santiago 763-0355, Chile}
\printaffiliation{jao}{Joint ALMA Observatory, Alonso de C\'ordova, 3107, Vitacura, Santiago 763-0355, Chile}
\printaffiliation{mpe}{Max Planck Institute for Extraterrestrial Physics, Gie{\ss}enbachstra{\ss}se 1, D-85748 Garching bei M{\"u}nchen, Germany}
\printaffiliation{ulaserena_postgrad}{Instituto Multidisciplinario de Investigaci\'on y Postgrado, Universidad de La Serena, Ra\'ul Bitr\'an 1305, La Serena, Chile}
\printaffiliation{ulaserena}{Departamento de Astronom\'ia, Universidad de La Serena, Ra\'ul Bitr\'an 1305, La Serena, Chile}
\printaffiliation{ice_csic}{Institut de Ci\`encies de l'Espai (ICE, CSIC), Campus UAB, Carrer de Can Magrans s/n, 08193, Bellaterra (Barcelona), Spain}
\printaffiliation{ieec}{Institut d'Estudis Espacials de Catalunya (IEEC), 08860 Castelldefels (Barcelona), Spain}
\printaffiliation{gbo}{Green Bank Observatory, 155 Observatory Road, Green Bank, WV 24944, USA}
\printaffiliation{utokyo}{Institute of Astronomy, The University of Tokyo, Mitaka, Tokyo 181-0015, Japan}
\printaffiliation{iaa_csic}{Instituto de Astrof\'{i}sica de
Andaluc\'{i}a, CSIC, Glorieta de la Astronomía s/n, 18008 Granada,
Spain}
\printaffiliation{manchester}{Jodrell Bank Center for Astrophysics, School of Physics and Astronomy, University of Manchester, Oxford road,  M13 9PL, United Kingdom.}

\bsp
\label{lastpage}
\end{document}